\documentclass[10pt,prd,aps,onecolumn,notitlepage,nofootinbib,nobalancelastpage,superscriptaddress]{revtex4-1}

\usepackage{amsmath} % Equations
\usepackage{amssymb} % Equations
\usepackage{ifthen} % for conditional statements
\usepackage{enumerate}
\usepackage{mciteplus}
\usepackage{tikz}
\usetikzlibrary{arrows.meta}
\usetikzlibrary{decorations.pathreplacing,calligraphy}
\usetikzlibrary{calc}
\usepackage{adjustbox}
\usepackage{bm}
\usepackage[colorlinks=true,
            urlcolor=blue,
            citecolor=blue,
            linkcolor=red,
            menucolor=blue,
            linktocpage=true]{hyperref}
\usepackage{graphicx}

\newboolean{pdflatex}
\setboolean{pdflatex}{false} % False for eps figures 
\newboolean{articletitles}
\setboolean{articletitles}{true} % False removes titles in references
\newboolean{uprightparticles}
\setboolean{uprightparticles}{True} %True for upright particle symbols

\def\mbf#1{\mathbf{#1}}
\def\mrm#1{\mathrm{#1}}
\def\be{\begin{equation}}\def\ee{\end{equation}}
\def\bea#1\eea{\begin{align}#1\end{align}}
\def\ssp{\hspace{0.09em}}
\def\sp{\hspace{0.06em}}
\def\Mid{\sp|\sp}

\def\rme{\mathrm{e}}
\def\im{\mathrm{i}}
\def\dif{\mathrm{d}}
\def\delD{\delta_\mathrm{D}}
\def\n{\hat{\mbf{n}}}
\def\calH{\mathcal{H}}
\def\x{\mbf{x}}
\def\s{\mbf{s}}
\def\r{\mbf{r}}

\def\k{\mbf{k}}
\def\q{\mbf{q}}
\def\v{\mbf{v}}

\def\u{\mbf{u}}

\def\J{\mbf{J}}
\def\ex{\mbf{e}_{x}}

\def\ez{\mbf{e}_{z}}
\def\U{\mbf{U}}
\def\T{\mathsf{T}}
\def\bA{b_\mrm{A}}
\def\bB{b_\mrm{B}}

\def\llangle{\langle}
\def\rrangle{\rangle_\delta}

\begin{document}
\title{Nonlinear Redshift-Space Distortions on the Full Sky}

\author{Lawrence Dam}
\email{lawrence.dam@unige.ch}
\affiliation{
D\'{e}partement de Physique Th\'{e}orique and
Centre for Astroparticle Physics,
Universit\'{e} de Gen\`{e}ve,
24 quai Ernest-Ansermet, CH-1211 Gen\`{e}ve 4, Switzerland}
\author{Camille Bonvin}
\email{camille.bonvin@unige.ch}
\affiliation{
D\'{e}partement de Physique Th\'{e}orique and
Centre for Astroparticle Physics,
Universit\'{e} de Gen\`{e}ve,
24 quai Ernest-Ansermet, CH-1211 Gen\`{e}ve 4, Switzerland}

\begin{abstract}

We derive an analytic expression for the two-point correlation function
in redshift space which (i) is nonlinear; ({ii}) is valid on the full sky,
i.e.\ the distant-observer limit is not assumed; (iii) can account
for the effect of  magnification and evolution bias due to a non-uniform
selection function; and (iv) respects the fact that observations are made on
the past lightcone, so naturally yields unequal-time correlations.
Our model is based on an exact treatment of the streaming model in the
wide-angle regime. Within this general regime, we find that the
redshift-space correlation function is essentially determined by a
geometric average of its real-space counterpart.
We show that the linear expression for the galaxy overdensity,
accurate to subleading order, can be recovered from our nonlinear framework.
This work is particularly relevant for the modeling of odd multipoles of
the correlation function at small separations and low redshifts,
where wide-angle effects, selection effects, and nonlinearities are
expected to be equally important.
\end{abstract}

\date{\today}
\maketitle

\section{Introduction}
Redshift-space distortions (RSD) have been identified as a key observable to test the laws of gravity and probe the validity of the $\Lambda$CDM model~\cite{Baker:2021}.
Typically one treats RSD in the flat-sky regime,
or distant-observer limit. In this regime, the redshift-space correlation function takes a simple form with a multipole structure consisting of a monopole, quadrupole and hexadecapole~\cite{Kaiser:1987qv,Hamilton:1992zz}. However, this approximation is only valid over a limited range of separations and opening angles. At small separations, nonlinearities become relevant and need to be included in the modeling; whereas at large separations and opening angles, the flat-sky approximation breaks down and wide-angle effects need to be accounted for. These two types of corrections are usually treated separately: either one models the linear correlation function with wide-angle corrections, or one models the flat-sky correlation function in the nonlinear regime. In most cases, these separate approaches are enough to provide a precise description of the signal.

Besides a desire for a general model,
there are two situations where both wide-angle corrections and nonlinear effects might become relevant over the same range of scales.
The first concerns measurements of the correlation function at very low redshift, such as those expected from the Dark Energy Spectroscopic Instrument (DESI)~\cite{DESI:2016fyo}. In particular, the Bright Galaxy Survey sample of DESI has a very high number density of galaxies at low redshift (with median $z\approx0.2$), over 14,000~$\text{deg}^2$~\cite{hahn2022desi}. At these redshifts, nonlinear evolution might be expected to be relevant up to relatively large separations, while wide-angle effects are expected to be important down to relatively small separations~\cite{Raccanelli:2010hk}. These effects are indeed governed by the ratio of pair separation $s$ to distance $d$, which quickly becomes non-negligible for small $d$. A nonlinear model on the full sky may therefore be needed for this type of survey.

The second situation concerns the measurement of relativistic effects~\cite{Yoo:2009au,Bonvin2011,Challinor:2011bk,Jeong:2011as}, where wide-angle effects and nonlinearities are both important over the same range of scales. Relativistic effects have been shown to contribute to the correlation function by generating odd multipoles (a dipole and an octupole) in the correlation of two populations of galaxies~\cite{McDonald:2009ud,Bonvin:2013ogt,Bonvin:2014owa}. Both in the linear regime~\cite{Bonvin:2013ogt} and in the perturbative nonlinear regime, wide-angle effects are roughly of the same order of magnitude as relativistic effects. This is because relativistic effects scale as $\mathcal{H}/k\times\text{RSD}$, while wide-angle effects scale as $s/d\times\text{RSD}$~\cite{Bonvin:2013ogt}. These two types of effects are therefore roughly of the same order of magnitude at all scales, since $s/d\sim \mathcal{H}/k$.\footnote{This is of course a crude comparison, since on the one hand the ratio of $\calH$ and $d$ varies with redshift, and on the other hand RSD and relativistic effects are also redshift-dependent. However, it shows that wide-angle effects and relativistic effects have a similar scaling and cannot be treated separately, even for large $k$.}
As a consequence, if one wants, for example, to measure through the dipole the relativistic effects in the nonlinear regime, it is necessary to model at the same time wide-angle effects in this regime.

A number of works have studied various aspects of the problem.
Castorina and White~\cite{Castorina_2018} calculated the impact of wide-angle corrections on the even multipoles, modelled using the
resummed approach to Lagrangian perturbation theory 
(LPT)~\cite{Taylor_1996,Matsubara:2008wx}. Their work showed that linear theory
is adequate to describe wide-angle corrections for the even multipoles,
except around the baryon acoustic peak where non-perturbative corrections
are known to be important~\cite{Baldauf:2015xfa}.
However, their model misses a contribution related to the (uniform)
selection function. This was pointed out by Taruya et al.~\cite{Taruya_2019},
who presented a model similar to that of Castorina and White, but without this
deficiency (see also Refs.~\cite{Saga:2020tqb,Saga:2021jrh} for subsequent work 
including the gravitational redshift). A comparison with simulations showed
that the contribution from the selection function is important for an accurate
prediction of the dipole moment (though not for the even multipoles).
Both of these works did not consider contributions from a \emph{non}-uniform radial 
selection function. It is however known that in the linear regime a non-uniform selection function generates contributions from the magnification bias,
which are of the same order of magnitude as wide-angle effects~\cite{Tansella:2017rpi}, and may even dominate the signal for some choices of populations~\cite{Bonvin:2023jjq}.

Concerning the second situation, Beutler and di Dio~\cite{Beutler:2020evf,DiDio:2020jvo} proposed a method to compute the relativistic power spectrum, including selection effects and wide-angle effects in perturbation theory. They derived an expression for the dipole, including contributions up to third order in perturbation theory, which agrees well with numerical simulations up to $k_{\rm max}\simeq 0.4\, h^{-1}$Mpc.
More recently, Noorikuhani and Scoccimarro~\cite{Noorikuhani_2023} calculated
the impact of relativistic
effects and wide-angle corrections on the galaxy power spectrum and bispectrum.
Their approach was to model these Fourier statistics in the usual way---i.e.\ work in
the distant-observer limit and use one-loop perturbation theory---but
supplement with the {leading-order} relativistic and wide-angle contributions.
This hybrid approach was justified on the basis
that the nonlinearities were found not to mix significantly with the 
relativistic and wide-angle effects.

This paper begins a study of these two situations from an altogether
different approach.
Here we will largely focus on the first situation, exhibiting
a novel approach to the streaming model~\cite{Scoccimarro_2004,Peebles_LSS},
a nonlinear model of the RSD correlation function; a forthcoming work
will be dedicated to a complete model for the second situation.
We will thus show how the streaming model can be exactly extended
to the wide-angle regime, taking advantage of the simple geometry of
the problem in configuration space.
This model is similar in some respects to that of
Taruya et al.~\cite{Taruya_2019} but differs importantly in others.
In particular, here we allow for the more
realistic case of a non-uniform selection function, which leads to further
distortions through the magnification and evolution bias. In addition,
here we derive in full generality the relation between the
matter density field in redshift space and in real space, independent of
the details of how such fields might evolve or might be biased in relation to
the galaxy field. (The dynamics and galaxy bias can be specified,
for example, using the `convolution LPT' 
prescription~\cite{Reid_2011,Carlson_2012,Wang_2013}, as has proven a
powerful method.)

Based on a more general treatment of the redshift mapping and number
conservation, we will further show that the streaming model can
also accommodate selection, galaxy evolution and relativistic
effects---indeed, almost all subleading effects at $\mathcal{O}(\calH/k)$.
These effects, as mentioned, are of the same order as the
wide-angle effects so in principle should also be taken into account.
With the streaming model, these effects are logically separated and
enter in resummed form, thereby
offering a compact way of capturing the large number of terms that
contribute to the overdensity at subleading order (i.e.\ when expressed 
through a  perturbative expansion). Additionally, its modular form
lends itself well to the problem of modeling at the same time
the three different kinds of sources of nonlinearity that need to
be considered in a realistic model---namely, dynamics, galaxy bias,
and the redshift mapping (wide-angle effects in our model are exact
to all orders in $s/d$).

The outline of this paper is as follows.
In Section~\ref{sec:wide} we extend the
nonlinear approach to RSD~\cite{Scoccimarro_2004} to the wide-angle
regime, deriving a non-perturbative expression for the
wide-angle correlation function in redshift space.
In Section~\ref{sec:fullmodel} we extend the derivation to
construct a more realistic model of the correlation function which takes into
account selection effects, as well as the fact that observations
are made on the lightcone.
We then perform a perturbative expansion of our model in Section~\ref{sec:recover-lt}
and show that
well-known results from linear theory are recovered, including of many relativistic effects.
In Section~\ref{sec:gaussian_stream} we present the full-sky generalisation
of the Gaussian streaming model, and show
that it is consistent with the expected form in the distant-observer limit.
In Section~\ref{sec:multipoles} we calculate the linear theory
multipoles including wide-angle contributions, and show that they are
consistent with expressions found in the literature.
Our conclusions follow in Section~\ref{sec:conclusions}.
Several appendices describe the details of our calculations.

\section{Nonlinear modeling in the wide-angle regime}\label{sec:wide}

This section is principally devoted to a study of the
relation between a galaxy at its true (comoving) position $\x$ and 
its observed (comoving) position $\s$,
\begin{align}\label{eq:sx}
    \s(\x)=\x+\hat\x\cdot\u(\x)\sp\hat\x \, ,
\end{align}
as concerns the correlation function in redshift space.
Here $\u=\calH^{-1}\v$ (which has units of length),
$\v$ is the peculiar velocity and $\calH$ is the conformal 
Hubble parameter. This mapping of course leads to the well-known
Kaiser effect, typically modelled in the distant-observer limit in which one
assumes that distant objects have identical line of sight $\hat\x$.
This approximation is valid for small opening angles
between any two lines of sight in a galaxy sample.

Here we present an exact treatment of the general case in which lines of
sight $\hat\x$ are allowed to vary across the full sky
without restriction to small angles.
To highlight the key trick in this paper and make clear the geometric
interpretation, we will first present the calculation of the full-sky
correlation function without any complicating factors
such as selection effects.
We will also focus on equal-time correlations and suppress the time
dependence in the number density, velocity, etc; we will restore it in Section 
\ref{sec:fullmodel} when we come to consider unequal-time correlations on
the lightcone and related projection effects.

The basis of our approach is the number conservation of objects in
real and redshift space:
\begin{align}
\label{eq:conservation}
n_s(\s)\ssp\dif^3\s=n(\x)\ssp\dif^3\x \, , 
\end{align}
where $n(\x)$ and $n_s(\s)$ are the comoving number densities
in real and redshift space, respectively.
In integral form, we have equivalently
\be\label{eq:ns}
n_s(\s)
=\int\dif^3\x\,
n(\x)\ssp\delD(\s-\s(\x)) \, .
\ee
This expression in fact holds for general mappings $\s(\x)$---not
just for Eq.~\eqref{eq:sx}---including those that also
contain transverse displacements. It also holds in the regime of
multiple streams, i.e.\ when more than one point in real space
formally maps to a single point in redshift space (when $\s(\x)$ has singular Jacobian).

Now, since we work on the full sky and since the mapping
only affects the radial positions, it is natural to switch
to a spherical coordinate system.
Thus let $\chi=|\sp\s\sp|$ be the (observed) radial distance in redshift space
and $\chi'=|\sp\x\sp|$ the radial distance in real space. Writing
Eq.~\eqref{eq:ns} in spherical coordinates, separating the Dirac delta
function into a radial piece and an angular piece, and inserting
$n_s(\s)=\bar{n}_s[1+\delta_s(\s)]$ and $n(\x)=\bar{n}[1+\delta(\x)]$ (here $\bar{n}_s$ and $\bar{n}$ denote the mean number densities in redshift and real space, respectively), we have
\bea
1+\delta_s(\s)
&=\int^\infty_0\dif\chi'\chi'^{\sp2}\int\dif^2\hat\x\,[1+\delta(\x)]\sp
    \frac{1}{\chi^2}\ssp\delD\big(\chi-\chi'-\hat\x\cdot\u(\x)\big)\ssp
    \delD(\n-\hat\x) \, ,
\eea
where $\n=\s/|\sp\s\sp|$ is the line of sight, and we have furthermore
used that, in the absence of selection or evolution effects, the
mean densities in real and redshift space are equal,
$\bar{n}=\bar{n}_s$; see Appendix~\ref{app:ns} for justification. 
Parametrising the positions as $\s=\chi\sp\n$ and $\x=\chi'\hat\x$,
and doing the trivial angular integral, we get
\bea
1+\delta_s(\chi\sp\n)
&=\frac{1}{\chi^2}\int^\infty_0\dif\chi'\chi'^{\sp2}\ssp
    [1+\delta(\chi'\n)]\ssp
    \delD(\chi-\chi'-\n\cdot\u(\chi'\n))
    \label{eq:dels-los-int-dirac}\\
&=\frac{1}{\chi^2}\int^\infty_0\dif\chi'\chi'^{\sp2}\ssp 
    [1+\delta(\chi'\n)]
    \int^\infty_{-\infty}\frac{\dif k}{2\pi}\;
    \rme^{-\im k(\chi-\chi')}\sp
    \rme^{\im k u_\|(\chi'\n)} \, ,
    \label{eq:dels-los-int-uniform}
\eea
where in the second line
the Dirac delta function is given as its Fourier representation,
writing $u_\|=\n\cdot\u$ for the radial component of the
velocity.
As we will show in Section~\ref{sec:recover-lt}, 
this equation recovers at linear order
the familiar Kaiser term, including the subdominant
inverse-distance term.
With Eq.~\eqref{eq:dels-los-int-uniform} it is straightforward to
obtain the correlation function
$\xi_s=\langle\delta_s(\s_1)\delta_s(\s_2)\rangle$:
% $\xi_s(\chi_1,\chi_2,\n_1\cdot\n_2)
% \equiv\langle\delta_s(\chi_1\n_1)\sp\delta_s(\chi_2\n_2)\rangle$:
\bea
1+\xi_s(\chi_1,\chi_2,\n_1\cdot\n_2)
&=\frac{1}{\chi_1^2\sp\chi_2^2}
    \int\dif\chi_1'\,\chi_1'^{\sp2}\int\dif\chi_2'\,\chi_2'^{\sp2}
    \int\frac{\dif^2\bm\kappa}{(2\pi)^2}\,
    \rme^{-\im\bm\kappa\cdot(\bm\chi-\bm\chi')}
    \big\langle[1+\delta(\chi_1'\n_1)][1+\delta(\chi_2'\n_2)]\ssp
        \rme^{\im\bm\kappa\cdot\mbf{w}}\big\rangle
    \label{eq:xis-2-alt} \\
&=\frac{1}{\chi_1^2\sp\chi_2^2}
    \int\dif\chi_1'\,\chi_1'^{\sp2}\int\dif\chi_2'\,\chi_2'^{\sp2}\,
    [1+\xi(r)]
    \int\frac{\dif^2\bm\kappa}{(2\pi)^2}\,
    \rme^{-\im\bm\kappa\cdot(\bm\chi-\bm\chi')}
    \llangle\sp\rme^{\im\bm\kappa\cdot\mbf{w}}\rrangle \, ,
    \label{eq:xis-2}
\eea
where $r=(\chi_1'^{\sp2}+\chi_2'^{\sp2}-2\chi_1'\chi_2'\cos\vartheta)^{1/2}$ 
is the separation between the two galaxies in real space,
$\cos\vartheta=\n_1\cdot\n_2$ is the cosine of the opening angle, and
we have defined the following two-component vectors:
$\bm\kappa=(k_1,k_2)$, $\bm\chi=(\chi_1,\chi_2)$,
$\bm\chi'=(\chi'_1,\chi'_2)$, and
$\mbf{w}\equiv(u_\|(\x_1),u_\|(\x_2))
=(\n_1\cdot\u(\chi_1'\n_1),\n_2\cdot\u(\chi_2'\n_2))$.
Furthermore, in the second line we have identified
the moment generator $\llangle\rme^{\im\bm\kappa\cdot\mbf{w}}\rrangle$,
where in this work a subscript $\delta$ denotes the
density-weighted ensemble average
\be\label{eq:density-weighted-avg}
\llangle O\rrangle
\equiv\langle[1+\delta(\x_1)][1+\delta(\x_2)]\sp O\sp\rangle
    \,/\, \langle[1+\delta(\x_1)][1+\delta(\x_2)]\rangle \, .
\ee

\begin{figure}[!t]
  \centering
  \includegraphics[scale=1]{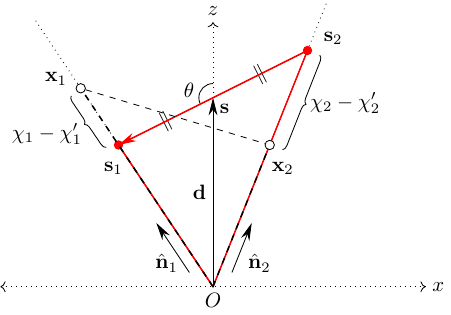}
  \caption{Parametrisation of a typical triangle configuration in the wide-angle regime.
The redshift-space configuration (indicated in red)
is the triangle formed by the
observer $O$ together with the galaxies at observed positions
$\s_1$ and $\s_2$. An example of a real-space configuration (dashed black)
that could map onto the redshift-space triangle
is indicated by the triangle formed by $\x_1$, $\x_2$, and $O$.
Note that candidates for the true positions, $\x_1$ and $\x_2$, can lie 
anywhere on the respective lines of sight, and the probability of a given
candidate triangle mapping onto the observed triangle is determined by 
the joint distribution of the separations
$\chi_1-\chi_1'$ and $\chi_2-\chi_2'$.
Here $\n_1$ and $\n_2$ are fixed, and we have aligned the $z$-axis to bisect
the separation $\s=\s_1-\s_2$ at the mid-point $\mbf{d}=(\s_1+\s_2)/2$.
Without loss of generality, the galaxy pair lives in the $xz$-plane.}
\label{fig:config}
\end{figure}

There is a more intuitive way of expressing Eq.~\eqref{eq:xis-2}.
Recognising that the $\bm\kappa$-integral in Eq.~\eqref{eq:xis-2}
(the inverse Fourier transform of the generating function) 
defines the joint probability distribution of radial displacements,
\be\label{eq:p-from-Z}
p(\bm\chi-\bm\chi';\bm\chi'\Mid\n_1\cdot\n_2)
\equiv\int\frac{\dif^2\bm\kappa}{(2\pi)^2}\,
    \rme^{-\im\bm\kappa\cdot(\bm\chi-\bm\chi')}\ssp
    Z(\J=\bm\kappa\sp;\r(\bm\chi'),\n_1\cdot\n_2) \, ,
\qquad
Z(\J;\r,\n_1\cdot\n_2)\equiv\llangle\rme^{\im\J\cdot\mbf{w}}\rrangle \, ,
\ee
we can write
\be
1+\xi_s(\chi_1,\chi_2,\n_1\cdot\n_2)
=\frac{1}{\chi_1^2\sp\chi_2^2}
    \int^\infty_0\dif\chi_1'\ssp\chi_1'^{\sp2}
    \int^\infty_0\dif\chi_2'\ssp\chi_2'^{\sp2}\ssp
    \big[1+\xi(r)\big]\ssp
    p(\bm\chi-\bm\chi';\bm\chi'\Mid\n_1\cdot\n_2) \, .
    \label{eq:xis-los-int}
\ee
This formula is the full-sky generalisation of the (distant-observer)
streaming model~\cite{Scoccimarro_2004}, which is given by a single
line-of-sight integral.

The probability distribution~\eqref{eq:p-from-Z} is scale-dependent:
it depends not only on $\bm\chi-\bm\chi'$ but {also} on $\bm\chi'$
itself through the moments of $\mbf{w}$ (by way of $Z$).\footnote{More 
precisely, these
velocity moments generally
depend on the separation $r$ and the
projections $\hat\r\cdot\n_1$ and $\hat\r\cdot\n_2$, which are geometrically
related to $\chi_1'$, $\chi_2'$ and $\n_1\cdot\n_2$.}
The existence of coherent flows is the origin of this scale
dependence, without which $p$ is a proper probability distribution.
This dependence on $\bm\chi'$ has the effect that as we integrate
over $\chi_1'$ and $\chi_2'$ we pass through a two-parameter family
of probability distributions, each with different mean, covariance,
etc---there is not a single fixed distribution.
There is also a dependence of $p$ on the opening angle $\n_1\cdot\n_2$
but, unlike the dependence on $\bm\chi'$, is known \emph{a priori}
(as indicated by the conditional).

A useful if heuristic way to view Eq.~\eqref{eq:xis-los-int} is
as the expectation of $1+\xi(r)$
when averaged over all real-space triangles that can be formed from an opening angle
$\n_1\cdot\n_2$, e.g.\ by varying the adjacent side lengths $\chi_1'$ and $\chi_2'$. 
Schematically, 
\be
\xi_s = \langle\xi\rangle_\Delta \, ,
\ee
for $\langle\cdot\rangle_\Delta$ some average over triangles.
More precisely, we have a probability space of real-space triangles, 
parametrised relative to the fixed redshift-space triangle by
$\chi_1-\chi_1'$ and $\chi_2-\chi_2'$ (see Fig.~\ref{fig:config}).
These radial displacements are correlated since they are the result
of Doppler shifts produced by the radial velocities
$u_\|(\chi_1'\sp\n)$ and $u_\|(\chi_2'\sp\n)$,
which are themselves correlated. Since velocity correlations
depend on the separation $\r=\x_1-\x_2=\chi_1'\n_1-\chi_2'\n_2$,
not all triangles in Eq.~\eqref{eq:xis-los-int} contribute with the same probability.
In particular, triangles in real space that are far from the redshift-space
configuration will contribute negligibly, since no large-scale correlated
flow is likely to arise that can map these configurations into each other;
conversely, configurations that are close to each other will contribute
significantly to the integral.
How close will depend on the characteristic separation along each line of
sight as determined by the means $\langle u_\|(\x_1)\rangle_\delta$ and
$\langle u_\|(\x_2)\rangle_\delta$.

We emphasise that no dynamical assumptions have been made in obtaining Eq.~\eqref{eq:xis-los-int};
it is an exact result based on the formal relation~\eqref{eq:ns} between the
observed and underlying density fields.%
\footnote{As with other nonlinear treatments of RSD
(e.g.~Ref.~\cite{Scoccimarro_2004}), our model is `exact' only to the
extent that the redshift mapping~\eqref{eq:sx} is exact.
But this {mapping} cannot be said to be exact as it is based on a
linear approximation of the full relation between $\s$ and $\x$
(even if the perturbations are themselves fully nonlinear);
see Section~\ref{sec:lookback}.
}
Furthermore, we have made no attempt to account for the galaxy bias,
since including it in this framework is 
straightforward~\cite{Desjacques_2018,Matsubara:2008wx,Carlson_2012}---e.g.~by 
replacing $1+\delta$ with $1+b\sp\delta$ (in linear theory), or, more generally,
some functional of $\delta$. Irrespective of tracer (galaxies, halos, dark matter particles, etc), the relation between the observed and underlying fields remains the same.

Finally, since the line-of-sight integrals in Eq.~\eqref{eq:xis-los-int}
are over non-oscillatory real functions, evaluating them numerically
is in principle straightforward once
the real-space correlation function and probability distribution
are specified.
(In Section~\ref{sec:gaussian_stream} we explicitly show the form of
these integrals in the case of the Gaussian streaming model.)

\section{A model including lightcone, selection and evolution effects}
\label{sec:fullmodel}

Going beyond the distant-observer limit, wide-angle effects are
among a number of effects that need to be considered all at once. 
We first give a physical explanation of these additional effects in Sections~\ref{sec:lookback} and~\ref{sec:selection}, 
and in Section~\ref{sec:derivemodel} we derive the full model including all effects.

\subsection{Extending the redshift mapping to the lightcone}
\label{sec:lookback}

Observations are made on the past lightcone but this is not reflected
in the mapping~\eqref{eq:sx} nor the correlation function~\eqref{eq:xis-los-int}
derived from it. In particular, the mapping~\eqref{eq:sx} does not take
into account the fact
that perturbations to the redshift also induce a displacement in the lookback time,
thus changing the apparent position on the lightcone.
We can see this by reconsidering the problem of mapping galaxies
in a redshift survey.%
\footnote{See Ref.~\cite{Jeong:2011as} for a discussion
of the general problem in terms of photon geodesics.}

The basic task is to assign Cartesian (comoving) coordinates $\s$ using
redshifts and angular positions.
For a galaxy with measured redshift $z$ observed in the direction
$\hat\x$, we assign 
to it the coordinates $\s=\chi(z)\sp\hat\x$, where the conversion
from redshift to comoving radial distance is given by
$\chi(z)=\int^z_0\dif z'/H(z')$. This is the observed position.
(Here we assume  perfect knowledge of the underlying background cosmology,
and no angular deflections so that $\hat\x=\n$.)
The (unknown) true position is $\x=\chi'\sp\hat\x$,
where $\chi'\equiv\chi(\bar{z})$, $\bar{z}=z-\delta z$ is the background
redshift and $\delta z$ the redshift perturbation. This is the position
that would be inferred had the redshift not suffered a Doppler shift.
The mapping~\eqref{eq:sx} is obtained by linearizing
$\s=\chi(\bar{z}+\delta z)\sp\hat\x$ about the true position
$\x=\chi(\bar{z})\sp\hat\x$:
\be\label{eq:sx-2}
\s
\simeq\chi(\bar{z})\ssp\hat\x 
    +\delta z\,\frac{\dif\chi}{\dif z}\bigg|_{z=\bar{z}}\,\hat\x
=\x+\hat\x\cdot\u(\tau',\x)\sp\hat\x \, ,
\ee
where in the second equality we have used that
$\delta z=(1+\bar{z})\ssp \v\cdot\hat\x$, obtained from the
relation $1+z=(1+\bar{z})(1+\v\cdot\hat\x)$ for the Doppler shift.

\begin{figure}[!t]
  \centering
  \includegraphics[scale=1]{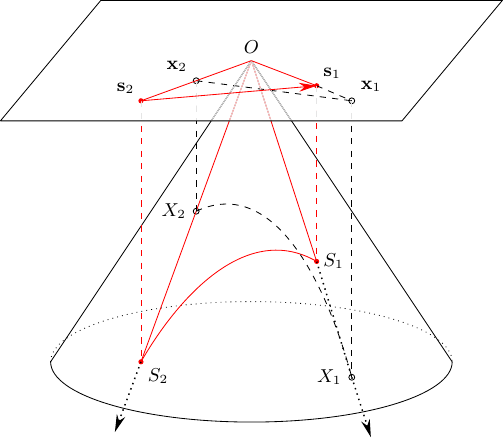}
  \caption{Configurations on the past lightcone and their spatial projections.
    The observed configuration is shown in red, while a potential true
    configuration is shown in black.
    Here $S_i=(\tau_0-\chi_i,\s_i)$ is the assigned spacetime position,
    while $X_i=(\tau_0-\chi_i',\x_i)$ is the real spacetime position, i.e.\ the
    position that would be observed absent all distortions to the redshift.
    The two-point correlation between $S_1$ and $S_2$ is determined by all
    two-point correlations between possible $X_1$ and $X_2$ falling on the
    lines of sight (dotted lines).}
\label{fig:lightcone-config}
\end{figure}

But redshift is not only an indicator of distance; it is also an
indicator of time, with galaxies at larger redshifts associated
with larger {lookback times}; see Fig.~\ref{fig:lightcone-config}.
So in addition to assigning spatial coordinates, we also assign
a time coordinate $\tau$ to each galaxy~\cite{Bonvin:2013ogt}.
More precisely, following
a galaxy photon along the line of sight $\hat\x$ back to the observed 
redshift $z$, we assign
\be\label{eq:lookback}
\tau=\tau_0-\chi(z) \, ,
\ee
the time at which the photon was apparently emitted.
Here $\tau_0$ is the present conformal time.
Likewise, we have for the real time $\tau'=\tau_0-\chi'$.
The relation between $\tau'$ and $\tau$ is then given by linearizing
$\tau(z)=\tau(\bar{z}+\delta z)$ about $\tau(\bar{z})\equiv\tau'$.
At linear order we have $\tau\simeq\tau'-u_\|(\tau',\x)$,
which together with Eq.~\eqref{eq:sx} constitutes a map
$(\tau',\x)\mapsto(\tau,\s)$ on the lightcone. Note that $\tau'$ is in fact
degenerate with $\x$ since $\tau'=\tau_0-\chi'$ and $\chi'=|\x|$, and that
the displacement on the lightcone is null: $-(\tau-\tau')^2+(\chi-\chi')^2=0$.

In the rest of this section we will work with this spacetime mapping.
As we will see in Section~\ref{sec:derivemodel}, any evolution in the number
density between the surface of constant $\tau$ (constant observed redshift $z$) and the surface of constant $\tau'$ (constant background redshift $\bar z$) gives rise to an apparent density fluctuation. These distortions are among
some of the many contributions to the full expression for the overdensity
derived using relativistic perturbation theory.
While subdominant to the Kaiser effect, these projection effects are of
the same order as wide-angle effects so in principle should also be included.

\subsection{Selection effects}
\label{sec:selection}

In addition to projection effects related to the lightcone, we also need
to take into account the selection effects.
These give rise to fluctuations of order $\calH/k$ which,
although subdominant to the usual RSD, are of the same size as
the wide-angle corrections so cannot generally be ignored.

\subsubsection{Flux limit}
Since surveys only observe above a certain flux limit $F_*$,
not all sources in the sky will be bright enough to be detected.
This is seen in the observed mean number density, or selection function 
$\bar{n}_s(\chi,F>F_*)$, which tends to fall off with distance $\chi$.
In general, we do \emph{not} have $\bar{n}_s=\bar{n}$ (where $\bar{n}$ is the
selection function in real space), since some sources that would otherwise not be detectable in real space, can be seen in redshift-space due to magnification effects (and vice-versa). The
difference between the two effectively generates an additional density 
fluctuation. Note that the selection function provides a
complete description of the selection effect in the nonlinear regime;
however in linear theory the relevant quantity is the linear response
of the selection function to a change in the flux limit, i.e.\ the slope of
$\bar{n}_s$ with respect to the threshold:
\be\label{eq:s_bias}
s_*\equiv \frac{\partial\log\bar{n}_s}{\partial\sp m_*}
=-\frac25\frac{\partial\ln\bar{n}_s}{\partial\ln F_*}
=-\frac25\frac{\partial\ln\bar{n}_s}{\partial\ln L_*}\, ,
\ee
where $m_*=-2.5\log F_*+\mrm{const.}$ is the magnitude limit
and $F_*$  ($L_*$) is the flux (luminosity) limit of the survey.
This parameter (not to be confused with the redshift coordinates) is
known as the `magnification bias' and is survey and population dependent.

\subsubsection{Galaxy evolution}
Galaxies can merge with each other or be created altogether. This was not reflected in the model~\eqref{eq:xis-los-int} which assumed a constant number of galaxies
($\bar{n}_s=\bar{n}=\text{const}$).
Since the time evolution of the mean comoving number density
$\bar{n}(\tau)$ depends upon the uncertain details of galaxy
formation and evolution, it is conventionally parametrised
by the `evolution bias'
\be\label{eq:fevol}
f_\mrm{evol} \equiv \frac{\partial\ln\bar{n}}{\partial\ln a} \; ;
\qquad
\bar{n}(\tau)
=F_\mrm{evol}(\tau)\ssp\bar{n}(\tau_0) \, ,
\qquad%\text{with}\qquad
F_\mrm{evol}(\tau)
\equiv
\exp\Big(-\int^1_{a(\tau)}\frac{\dif a'}{a'}\, f_\mrm{evol}(a')\Big) \, .
\ee
With no evolution, $f_\mrm{evol}=0$ and $F_\mrm{evol}=1$ for all time.
In general, $f_\mrm{evol}$ is tracer dependent and a function of
the flux cut.
Note that the effect of galaxy evolution on the apparent number density
may be considered an example of a projection effect, in that
the lookback time~\eqref{eq:lookback} of a galaxy viewed in real space
is different to the lookback time of the same galaxy but viewed in
redshift space.

\subsection{Derivation of the general model}
\label{sec:derivemodel}

We now construct
a model of the redshift-space correlation function
valid on the full sky and in the nonlinear regime, building into it
the lookback time~\eqref{eq:lookback}, as well as
the flux cut and galaxy evolution.
The calculation is essentially the same as before once we have
setup the problem and introduced some definitions.
Readers who are not interested
in these details may skip ahead to Eq.~\eqref{eq:dels-los-int}
and follow the discussion from there.

To include a flux cut in the model we now need to consider the
luminosity of each galaxy. 
We define the redshift-space distribution function 
$\Phi_s(\tau,\s,F_s)$, i.e.\ 
the redshift-space comoving number density of galaxies
in the flux bin $(F_s,F_s+\dif F_s)$.
Similarly, let $\Phi(\tau,\x,F)$ be the true distribution function,
i.e.\ the real-space comoving number density
of galaxies in the (non-redshifted) flux bin $(F,F+\dif F)$.

Since the mapping $(\tau',\x)\mapsto(\tau(\x),\s(\x))$,
where $\tau(\x)$ is the lookback time~\eqref{eq:lookback} and $\s(\x)$ is
given by Eq.~\eqref{eq:sx}, is nothing more than a reassignment of each
galaxy's coordinates, the number of galaxies per flux bin is conserved:
\be\label{eq:conservation-lum}
\Phi_s(\tau,\s,F_s)\ssp\dif^3\s\,\dif F_s
=\Phi(\tau',\x,F)\ssp\dif^3\x\,\dif F
=\Phi(\tau',\x,L)\ssp\dif^3\x\,\dif L\, ,
\ee
where $\tau'=\tau_0-\chi'$ and, in a slight abuse of notation,
$\Phi(\tau',\x,F)=\Phi(\tau',\x,L)\,\dif L/\dif F$.
Here $\tau=\tau_0-\chi$ and $\tau'=\tau_0-\chi'$, and we
recall that these are related by $\tau=\tau'-u_\|$.
Equation~\eqref{eq:conservation-lum} simply reflects the fact that all galaxies observed in $\dif^3\s$ with flux between $F_s$ and $F_s+\dif F_s$, physically lie in $\dif^3\x$ with intrinsic luminosity between $L$ and $L+\dif L$.
Among all the galaxies in the volume element $\dif^3\s$ or $\dif^3\x$,
we select only those that meet or exceed the flux threshold $F_*$:
\be\label{eq:threshold-phi-numcon}
\Theta(F_s-F_*)\sp\Phi_s(\tau,\s,F_s)\ssp\dif^3\s\,\dif F_s
=\Theta(L-L_*(\x))\sp\Phi(\tau',\x,L)\ssp\dif^3\x\,\dif L \, .
\ee
Here $\Theta$ is the Heaviside step function which
enforces the threshold and
$L_*(\x)=4\pi\sp d_L^{\sp2}(\x)\sp F_*$ is the luminosity threshold
for an object at luminosity distance $d_L(\x)$.
Note that on both sides of Eq.~\eqref{eq:threshold-phi-numcon} we are
imposing the same selection criterion so that the same galaxies
are being selected in both real and redshift space. Since the luminosity distance $d_L$ is affected by inhomogeneities and depends therefore on direction, a fixed flux threshold $F_*$ in all directions corresponds to different luminosity thresholds $L_*(\x)$ in different directions.

Integrating both sides of Eq.~\eqref{eq:threshold-phi-numcon}
yields the differential relation between number densities 
[cf.~Eq.~\eqref{eq:conservation}]
\begin{align}
\label{eq:conservation_flux}
n_s(\tau,\s;F>F_*)\ssp\dif^3\s=n(\tau',\x; L>L_*(\x))\ssp\dif^3\x \, ,
\end{align}
where
\bea
n_s(\tau,\s;F>F_*)
\equiv\int^\infty_{F_*}\dif F_s\,\Phi_s(\tau,\s,F_s) \, ,
\qquad
n(\tau',\x;L>L_*(\x))
\equiv\int^\infty_{L_*(\x)}\dif L\,\Phi(\tau',\x,L) \, .
\eea

Separating the number densities into a mean contribution and an overdensity, 
assuming a universal luminosity function, we obtain
\begin{align}
\label{eq:deltas_gen}
[1+\delta_s(\tau,\s)]\ssp\dif^3\s
=\frac{\bar{n}(\tau',L>L_*(\x))}{\bar{n}_s(\tau,F>F_*)}\ssp
    [1+\delta(\tau',\x)]\ssp\dif^3\x \, .
\end{align}
The denominator on the right-hand side can be rewritten as
\begin{align}
\bar{n}_s(\tau,F>F_*)=\bar{n}(\tau, L>\bar{L}_*(\chi))\, ,
\end{align}
since the mean number of galaxies at $\tau$ with flux above $F_*$ corresponds to the galaxies that have mean intrinsic luminosity above $\bar{L}_*(\chi)$. The fraction in Eq.~\eqref{eq:deltas_gen} can then be split as
\begin{align}
\frac{\bar{n}(\tau',L>L_*(\x))}{\bar{n}(\tau, L>\bar{L}_*(\chi))} =\frac{\bar{n}(\tau',L>L_*(\x))}{\bar{n}(\tau', L>\bar{L}_*(\chi'))}
\frac{\bar{n}(\tau', L>\bar{L}_*(\chi'))}{\bar{n}(\tau, L>\bar{L}_*(\chi'))}\frac{\bar{n}(\tau, L>\bar{L}_*(\chi'))}{\bar{n}(\tau, L>\bar{L}_*(\chi))}\, ,
\end{align}
which gives rise to three contributions. 
First $\delta_*$, defined as
\begin{align}
\label{eq:deltastar}
1+ \delta_*(\tau',\x)=\frac{\bar{n}(\tau',L>L_*(\x))}{\bar{n}(\tau', L>\bar{L}_*(\chi'))}\, ,   
\end{align}
represents the fractional number density of galaxies at $\x$ with luminosity comprised between $\bar{L}_*(\chi')$ and $L_*(\x)=\bar{L}_*(\chi')+\delta L_*(\chi',\n)$, where the
perturbation to the luminosity threshold $\delta L_*$ is directly
related to the perturbation to the luminosity distance by
\begin{align}
\label{eq:deltalum}
\delta L_*(\chi',\n)
% =L_*(\x) - \bar{L}_*(\chi')
=4\pi F_* \big[d_L^{\sp2}(\chi',\n)-\bar{d}_L^{\,2}(\chi')\big]\, ,
\end{align}
and is affected by the Doppler effect (among other things).
Here $\bar{L}_*(\chi')=4\pi \bar{d}_L^{\,2}(\chi')F_*$ is the threshold
that would be adopted in the absence of perturbations to the
luminosity distance. 

Second $\delta_{\rm evol}$, defined as
\begin{align}
1+\delta_{\rm evol}(\tau',\x)=\frac{\bar{n}(\tau', L>\bar{L}_*(\chi'))}{\bar{n}(\tau, L>\bar{L}_*(\chi'))} \, ,   
\end{align}
encodes the evolution of the comoving number density of galaxies, above a \emph{fixed} luminosity threshold $\bar{L}_*(\chi')$, between the hypersurface of constant $\tau$ (corresponding to constant observed redshift $z$) and the hypersurface of constant $\tau'$ (corresponding to constant background redshift $\bar{z}$). Using Eq.~\eqref{eq:fevol}, applied at $\tau$ and $\tau'$ for the same luminosity threshold $\bar{L}_*(\chi')$, we obtain
\begin{align}
\label{eq:deltaFevol}
1+\delta_{\rm evol}(\tau',\x)=\frac{F_{\rm evol}(\tau')}{F_{\rm evol}(\tau)} \, .   
\end{align}

Finally $\delta_L$, defined as 
\begin{align}
\label{eq:deltaL}
1+\delta_L(\tau',\x)=\frac{\bar{n}(\tau, L>\bar{L}_*(\chi'))}{\bar{n}(\tau, L>\bar{L}_*(\chi))} \, ,  
\end{align}
describes the fractional number of galaxies with luminosity between $\bar{L}_*(\chi')$ and $\bar{L}_*(\chi)$. (Here $\delta_L$ is not to be confused with the linear matter field.) This term accounts for the fact that, since we select galaxies above a fixed flux threshold $F_*$, we do not select the same population of galaxies at each distance. Galaxies that are further away are selected with a higher luminosity than galaxies that are closer. Because of this, even if the luminosity function would be constant in time for all values of $L$, there is an evolution in the mean number density.

With this, Eq.~\eqref{eq:deltas_gen} becomes
\begin{align}
[1+\delta_s(\tau,\s)]\ssp\dif^3\s=[1+\delta_{\rm tot}(\tau',\x)]\ssp\dif^3\x \, ,
\end{align} 
where
\begin{align}
\label{eq:deltatot}
1+\delta_{\rm tot}(\tau',\x)\equiv [1+\delta_{\rm evol}(\tau',\x)][1+\delta_L(\tau',\x)][1+\delta_*(\tau',\x)][1+\delta(\tau',\x)] \, .
\end{align} 
Without selection and evolution effects, clearly $\delta_\mrm{tot}=\delta$.

An explicit expression for $\delta_s$ can now be obtained by a similar 
calculation to the one presented in 
Section~\ref{sec:wide}. Thus, passing from differential to integral 
form~\eqref{eq:ns}, changing to spherical coordinates, etc, we have
\be
1+\delta_s(\tau,\chi\sp\n)
=\frac{1}{\chi^2}\int^\infty_0\dif\chi'\sp
    \chi'^{\sp2}\sp
    \big[1+\delta_{\rm tot}(\tau_0-\chi',\chi'\n)\big]\ssp
    \int\frac{\dif k}{2\pi}\;
    \rme^{-\im k(\chi-\chi')}
    \exp\big[\im\sp k\sp u_\|(\tau_0-\chi',\chi'\n)\big]\, .
    \label{eq:dels-los-int}
\ee
The difference between this expression and our earlier
one, Eq.~\eqref{eq:dels-los-int-uniform}, is that $\delta$ there is
replaced with $\delta_\mrm{tot}$ here, and secondly the line-of-sight integral
is now performed on the past lightcone.

Finally, since Eq.~\eqref{eq:dels-los-int} is of the same
form as Eq.~\eqref{eq:dels-los-int-uniform}, the calculation
proceeds as before and we can write down at once the correlation function
[cf.~Eq.~\eqref{eq:xis-los-int}]:
\bea
1+\xi_s(\chi_1,\chi_2,\n_1\cdot\n_2)
=   \frac{1}{\chi_1^2\sp\chi_2^2}
    \int^\infty_0\dif\chi_1'\:\chi_1'^{\sp2}
    \int^\infty_0\dif\chi'_2\:\chi_2'^{\sp2}\ssp
    \big[1+\xi_{\rm tot}(\chi_1',\chi_2',\n_1\cdot\n_2)\big]\ssp
    p_\mrm{tot}(\bm\chi-\bm\chi'\Mid\n_1\cdot\n_2) 
    \label{eq:xis-los-int-bias} \\[4pt]
\qquad\qquad\text{with}\quad
1+\xi_{\rm tot}(\chi_1',\chi_2',\n_1\cdot\n_2) \equiv
\big\langle[1+\delta_\mrm{tot}(\tau_0-\chi_1',\chi_1'\n_1)]\sp
        [1+\delta_\mrm{tot}(\tau_0-\chi_2',\chi_2'\n_2)]\big\rangle\, ,
\nonumber
\eea
where $p_\mrm{tot}$ is given by Eq.~\eqref{eq:p-from-Z}, in which
the density weighting~\eqref{eq:density-weighted-avg} is now
with respect to $\delta_{\rm tot}$ (instead of $\delta$).
% Here $1+\xi_{\rm tot} \equiv
% \langle[1+\delta_\mrm{tot}(\tau_1',\x_1)]\sp[1+\delta_\mrm{tot}(\tau_2',\x_2)]\rangle$, with $\tau_1'=\tau_0-\chi_1'$ and $\tau_2'=\tau_0-\chi_2'$.
% (Note that we do not necessarily have $\langle\delta_\mrm{tot}\rangle=0$.)
In particular, notice that $\xi_\mrm{tot}$ no longer depends on the separation 
$r$, as was the case before, but on the full triangular configuration
given by three numbers, namely the side lengths $\chi_1'$, $\chi_2'$, and the
opening angle $\n_1\cdot\n_2$. As we will see in Section~\ref{sec:recover-lt},
this is because in a perturbative
expansion $\delta_\mrm{tot}$ contains terms depending on the line of sight
$\n$, whose presence induces an angular dependence in the correlations,
breaking statistical isotropy.

We can also see that the correlation function~\eqref{eq:xis-los-int-bias}
is a function of the (apparent) past lightcone:
it manifestly expresses the unequal-time correlation of any two points
$S_1=(\tau_1,\s_1)=(\tau_0-\chi_1,\chi_1\n_1)$ and
$S_2=(\tau_2,\s_2)=(\tau_0-\chi_2,\chi_2\n_2)$.
As with our earlier Eq.~\eqref{eq:xis-los-int}, the formula we have just 
derived is still `summing over triangles', but now the `triangles' are all 
those configurations that can be formed on the past light cone from an
opening angle $\n_1\cdot\n_2$, as opposed to those formed on spatial hypersurfaces.
Since these configurations are on the lightcone, they are in principle
all observationally accessible, e.g.\ if the peculiar velocity of each
galaxy was perfectly known. 
Hence the correlation function~\eqref{eq:xis-los-int-bias}
is determined by marginalising over all potentially observable
configurations, unlike our earlier Eq.~\eqref{eq:xis-los-int}, which is 
determined by marginalising over unobservable configurations.

Finally, $\xi_s$ is expressed in terms of radial distances
(assuming perfect knowledge of the underlying background cosmology),
but we note that it is also possible, and perhaps more desirable, to
express it in terms of observed and background redshifts, $z$ and $\bar{z}$. 
Working in terms of redshifts and angles,
the natural observable is the three-dimensional angular power spectrum
$C_\ell(z,z')$~\cite{Bonvin2011,Tansella:2017rpi}, which we note can be
constructed from our $\xi_s$ if we leave arbitrary the conversion
of redshift to distance.%
\footnote{See Refs.~\cite{Raccanelli:2023a,Raccanelli:2023b,Gao:2023rmo}
for related work on connecting $C_\ell(z,z')$ to the idealised
power spectrum $P(k)$, corrected for unequal-time correlations and
wide-angle effects.}

\section{Recovering linear theory}\label{sec:recover-lt}
We now verify that our nonlinear expression~\eqref{eq:dels-los-int} of the
redshift-space overdensity recovers well-known results from linear theory.
To begin, we expand the last exponential in Eq.~\eqref{eq:dels-los-int}, keeping only up to the linear contribution in $u_\|$:
\bea
1+\delta_s(\tau,\chi\sp\n)
&\simeq\frac{1}{\chi^2}
    \int\dif\chi'\sp\chi'^{\sp2}\sp
    \big[1+\delta_{\rm tot}(\tau',\chi'\n)\big]
    \int\frac{\dif k}{2\pi}\,
    \rme^{-\im k(\chi-\chi')}
    \big(1+\im k\sp u_\|(\tau',\chi'\n)\big)
    \label{eq:dels-exp1}\\
&=\frac{1}{\chi^2}
    \int\dif\chi'\sp\chi'^{\sp2}\sp
    \big[1+\delta_{\rm tot}(\tau',\chi'\n)\big]\,
    \bigg(\delD(\chi-\chi')
        -u_\|(\tau',\chi'\n)\frac{\dif}{\dif\chi}\delD(\chi-\chi')
    \bigg) \, , \nonumber
\eea
where we have written factors of $k$ as radial derivatives using
$\im k\sp\rme^{-\im k\chi}=-\dif\sp\rme^{-\im k\chi}/\dif\chi$. Note that in Eq.~\eqref{eq:dels-exp1} we have made explicit the dependence of the field in the position $\x=\chi'\n$ but also in time $\tau'=\tau_0-\chi'$, since the density and velocity are evolving with time. Dropping the quadratic term $u_\|\sp\delta$, taking the $\chi$-derivative outside of the integral and doing the now trivial integrals, 
we find (recalling that $u_\|\equiv\calH^{-1}v_\|$)
\begin{align}
1+\delta_s(\tau,\chi\sp\n)
% &=1+\delta(\chi\sp\n)
%     -\frac{1}{\chi^2\ssp\bar{n}_s(\chi)}
%         \frac{\partial}{\partial\chi}
%         \int\dif\chi'\sp\chi'^{\sp2}\sp
%         \bar{n}_s(\chi') u_\|(\chi'\n)\ssp \delD(\chi-\chi')
%     \nonumber\\
&=1+\delta_{\rm tot}(\tau,\chi\sp\n)
    -\frac{1}{\chi^2}
        \frac{\dif}{\dif\chi}
        \bigg(\chi^2\sp\frac{v_\|(\tau,\chi\sp\n)}{\calH(\tau)}\bigg)
        \nonumber\\[2pt]
&=1+\delta_{\rm tot}(\tau,\chi\sp\n)
    -\frac{1}{\chi^2}
        \frac{\partial}{\partial\chi}
        \bigg(\chi^2\sp\frac{v_\|(\tau,\chi\sp\n)}{\calH(\tau)}\bigg)
        -\frac{1}{\chi^2}\frac{\dif\tau}{\dif\chi}
        \frac{\partial}{\partial\tau}
        \bigg(\chi^2\sp\frac{v_\|(\tau,\chi\sp\n)}{\calH(\tau)}\bigg)
        \nonumber\\[2pt]
&=1+\delta_{\rm tot}(\tau,\chi\sp\n)
-\frac{1}{\calH}\frac{\partial\sp v_\|}{\partial\chi}
-\bigg(\frac{2}{\calH\chi}+\frac{\dot{\calH}}{\calH^2}\bigg)\ssp v_\|
+\frac{1}{\calH}\sp\dot{v}_\| \, ,
    %-\frac{\partial}{\partial\chi}u_\|(\chi\sp\n)\
    %-\frac{2}{\chi}\ssp u_\|(\chi\sp\n).
    \label{eq:dels-lin1}
\end{align}
where $\tau=\tau_0-\chi$ and an overdot denotes
partial differentiation with respect to conformal time.
Note that the second and third term in the second line is equal to
$\delta_s=\delta_\mrm{tot}-\bm\nabla\cdot\u_\|$
(with the radial velocity field $\u_\|\equiv\n\cdot\u(\x)\sp\n$),
a formula that is conventionally obtained by linearizing the Jacobian
of the mapping~\eqref{eq:sx}. The last two terms in the last line of
Eq.~\eqref{eq:dels-lin1} arise because of the lookback time, and
are are thus not captured by the Jacobian.

The kinematic terms in Eq.~\eqref{eq:dels-lin1} can be understood as follows.
The third term, the radial derivative of the velocity, gives the well-known RSD effect. The fourth term, proportional to $2/(\calH\chi)$, is the wide-angle contribution from a uniform selection function, already present in the original Kaiser formula. This term is usually ignored since at large distances (compared to the
separation) it is subdominant to the standard RSD term, but it is important to include in a wide-angle analysis~\cite{Raccanelli_2010}. Less well known are the last two terms in Eq.~\eqref{eq:dels-lin1}; these are due to the fact that, because we are integrating along the line of sight, we are traversing
a geodesic on the past lightcone, with both $\calH$ and $v_\|$ evolving along it~\cite{Bonvin:2013ogt}.

In addition to these terms, which derive from the mapping itself,
there are also the selection and evolution effects. These are
contained in $\delta_\mrm{tot}$, Eq.~\eqref{eq:deltatot}, which
at linear order reads
$\delta_\mrm{tot}\simeq\delta+\delta_\mrm{evol}+\delta_L+\delta_*$.
We calculate $\delta_\mrm{evol}$, $\delta_L$ and $\delta_*$ as follows.
\begin{enumerate}[(i)]
\item
For $\delta_{\rm evol}$, expand
$F_\mrm{evol}(\tau')$
in Eq.~\eqref{eq:deltaFevol} about $\tau=\tau'+\delta\tau$:
\be\label{eq:taylorevol}
F_\mrm{evol}(\tau')\simeq F_\mrm{evol}(\tau)
    -\frac{\dif F_\mrm{evol}}{\dif\tau}\,\delta \tau
=F_\mrm{evol}(\tau)\sp
    \big(1-f_\mrm{evol}\ssp\calH\ssp \delta\tau\big)  \, ,
\ee
where in the second equality the derivative has been evaluated
using Eq.~\eqref{eq:fevol}.
Inserting this into Eq.~\eqref{eq:deltaFevol} and using that by
Eqs.~\eqref{eq:lookback} and \eqref{eq:sx}
$\delta\tau=-\delta\chi=-\calH^{-1}v_\|$, we find
\be\label{eq:delta_evol_lin}
\delta_\mrm{evol}=f_\mrm{evol}\ssp v_\| \, .
\ee
%%%%%%%%%%%%%%%%%%%%%%%%%%%%%%%
\item
For $\delta_L$, expand $\bar{n}(\tau,L>\bar{L}_*(\chi'))$
in Eq.~\eqref{eq:deltaL} around $\chi=\chi'+\delta\chi$.
At linear order,
\be\label{eq:nbar_chi}
\bar{n}(\tau,L>\bar{L}_*(\chi'))
\simeq \bar{n}(\tau,L>\bar{L}_*(\chi))\left(1
    -\frac{\partial\ln\bar{n}}{\partial\ln\bar{L}_*}
        \frac{\dif\ln\bar{L}_*}{\dif\ln\chi}\ssp
        \frac{\delta\chi}{\chi}\right) \, ,
\ee
% \begin{align}
% \label{eq:derivevol}
% \frac{d\bar{n}(\chi,L>\bar{L}_*(\chi))}{d\chi}=\frac{\partial\bar{n}(\chi,L>\bar{L}_*(\chi))}{\partial\chi} +\frac{\partial\bar{n}(\chi,L>\bar{L}_*(\chi))}{\partial L_*}\frac{d L_*(\chi)}{d\chi}\, .
% \end{align}
where all quantities on the right-hand side are evaluated at $\chi$
and the chain rule has been used on the second term. Here
$\delta\chi=\calH^{-1}\sp v_\|$ and the $\chi$ derivative is
\begin{align}
\frac{\dif\ln\bar{L}_*}{\dif\ln\chi}
=2\sp\frac{\dif\ln\bar{d}_L}{\dif\ln\chi}
=2\sp\calH\chi\sp\bigg(1+\frac{1}{\calH\chi} \bigg)\, ,
\label{eq:dL_star}
\end{align}
where the first equality follows because $\bar{L}_*\propto \bar{d}_L^{\,2}$
and the second equality follows from differentiating
$\bar{d}_L(\chi)=(1+z)\ssp\chi=\chi/a[\tau(\chi)]$.
Inserting the linear expansion~\eqref{eq:nbar_chi} into Eq.~\eqref{eq:deltaL}, we have
\begin{align}
\label{eq:deltaevolres}
\delta_L=5\sp s_*\sp\bigg(1+\frac{1}{\calH\chi} \bigg)\ssp v_\| \, ,
\end{align}
where we have inserted Eq.~\eqref{eq:s_bias} for the magnification bias,
replacing $\bar{n}$ with $\bar{n}_s$ (since the difference results in a
second-order correction).

%%%%%%%%%%%%%%%%%%%%%%%%%%%%%%%
\item
For $\delta_*$, expand $\bar{n}(\tau',L>L_*(\x))$ in Eq.~\eqref{eq:deltaL}
around $\bar{L}_*(\chi')=L_*(\x)-\delta L_*(\chi',\n)$. At linear order,
\bea\label{eq:nbar_*}
\bar{n}(\tau',L>L_*(\x))\simeq \bar{n}(\tau',L>\bar{L}_*(\chi'))\left(1+\frac{\partial \ln\bar{n}}{\partial \ln\bar{L}_*} \frac{\delta L_*(\chi',\n)}{\bar{L}_*(\chi')} \right)\, .
\eea
The perturbation to the threshold at a fixed position in real space is
\begin{align}
\frac{\delta L_*(\chi',\n)}{\bar{L}_*(\chi')}=2\sp\frac{\delta d_L(\chi',\n)}{\bar{d}_L(\chi')}=4\sp v_\|\, ,
\end{align}
where the first equality follows from linearizing Eq.~\eqref{eq:deltalum},
while in the second equality we have used the luminosity distance fluctuations due to the source velocity calculated in Ref.~\cite{Bonvin:2005ps}.% 
\footnote{See equation (53) (or equivalently equation (55)) in Ref.~\cite{Bonvin:2005ps}, noting that $\n$ there is equal to
$-\n$ here.}
Inserting the linear expansion~\eqref{eq:nbar_*} into Eq.~\eqref{eq:deltastar},
we obtain
\begin{align}
\label{eq:deltaLres}
\delta_*
% =4\ssp\frac{\partial\ln\bar{n}}{\partial\ln\bar{L}_*}\ssp v_\|
=-10\sp s_*\sp v_\|\, ,
\end{align}
where again we have substituted in Eq.~\eqref{eq:s_bias} for the
magnification bias.
% Note that in Eq.~\eqref{eq:taylorevol} there is a total derivative in the second term, since the luminosity distance $\bar{L}_*$ is a function of $\chi'$ in the left-hand side, that we want to evaluate at $\chi$ in the right-hand side.
%%%%%%%%%%%%%%%%%%%%%%%%%%%%%%%%%%%%%%
\end{enumerate}

Finally, inserting Eqs.~\eqref{eq:delta_evol_lin}, \eqref{eq:deltaevolres},  
and~\eqref{eq:deltaLres}
into Eq.~\eqref{eq:dels-lin1} for $\delta_\mrm{tot}$, we obtain
\begin{align}
\delta_s(\tau,\chi\sp\n)=\delta
-\frac{1}{\calH}\frac{\partial\sp v_\|}{\partial\chi}+\frac{1}{\calH}\ssp\dot{v}_\| + \left(f_{\rm evol}-5s_*-\frac{\dot{\calH}}{\calH^2}+\frac{5s_*-2}{\calH\chi}\right)v_\| \, . 
\label{eq:dels-lin-all}
\end{align}
Upon comparing this equation with the full expression obtained
from relativistic calculations---e.g.\ equation (2.13) in
Ref.~\cite{Bonvin:2020cxp}---we see that we have recovered all subleading effects at $\mathcal{O}(\calH/k)$, \emph{with the exception of two terms}.
The first is a kinematic term given simply as $v_\|$.
This missing term can be traced back to the starting point of our derivation, Eq.~\eqref{eq:conservation}, which is based on the naive Euclidean volume element $\dif^3\x$. This Newtonian derivation does not account for the fact that the hypersurface of constant time for the moving galaxies (in real space) does {not} coincide with the hypersurface of constant conformal time $\tau$. That is, these two frames are `tilted' with respect to one another, and it is by accounting for this that we recover
precisely the term that we are after.
From an observational point of view, this term arises because photons, followed
back down the past lightcone, do not probe the rest-frame
galaxy density. Based on purely kinematic considerations, these
photons will intercept more galaxies moving towards them versus away from them~\cite{Kaiser_2013}, so that if a galaxy is receding
away from the observer then the apparent local density is enhanced relative
to its intrinsic value. 
Technically speaking, the missing term arises through projection of the four-current $j^\mu=n u^\mu$ at the source position onto the covariant three-dimensional volume element (the three-form dual to the one-form $\dif x^\mu$). 
Clearly this requires a relativistic treatment, beginning with a covariant notion of number conservation~\cite{Dam:2023}; this is however
beyond the scope of this work. 

The second term missing is $\calH^{-1}\partial\Psi/\partial\chi$, the
contribution from the gravitational redshift. This can be put down to the simple fact that the standard mapping~\eqref{eq:sx} only accounts for the
dominant Doppler shift and therefore ignores the subdominant contribution from the gravitational redshift. 
To illustrate the basic structure in a minimal model, we have neglected
to include the gravitational redshift. However, adding this effect into the
model is straightforward: by Eq.~\eqref{eq:sx-2} we take
$\s(\x)\to\s(\x)-\calH^{-1}\Psi(\x)\hat\x$;
see also Refs.~\cite{Saga:2020tqb,Saga:2021jrh}.
A complete model including the gravitational redshift and the
relativistic tilt will be presented in a
forthcoming work~\cite{Dam:2023}.

\section{Gaussian streaming model on the full sky}
\label{sec:gaussian_stream}

The discussion up to now has been fairly general in that
no assumptions have been placed on the statistics of the velocities
that determine $p(\bm\chi-\bm\chi')$ and therefore the correlation
function.
We now wish to specify these statistics by presenting a particular model of 
Eq.~\eqref{eq:xis-los-int-bias}, namely,
the full-sky version of the Gaussian streaming model~\cite{Reid_2011},
often used in configuration-space analyses~\cite{Samushia_2014,BOSS:2016ntk,Zarrouk:2018,Bautista:2020}.
We will however include the selection and evolution effects,
which we recall are entirely contained in $\delta_\mrm{tot}$,
Eq.~\eqref{eq:deltatot}. The lookback time is also included, which
amounts to taking $\tau'\to\tau_0-\chi'$.

We follow the usual procedure~\cite{Scoccimarro_2004,Matsubara:2008wx} for
constructing such models, namely, we rewrite the
generating function $Z$ in terms of the connected moments using
the cumulant generating function $W\equiv\ln Z$,
% or $\langle\rme^{jX}\rangle_\mrm{c}=\ln\langle\rme^{jX}\rangle$
% (where subscript `c' denotes the connected part),
then Taylor expand $W$, keeping only the first and second connected moments
(as determines a Gaussian).
In detail, by expanding $W(\J)\equiv\ln Z(\J)$ about $\J=0$ we have
\be
W(\J)
=\sum_{n=1}^\infty\frac{\im^n}{n!}\,
    J_{i_1}\cdots J_{i_n}\,
    \langle w_{i_1}\cdots w_{i_n}\rangle_{\delta_\mrm{tot},\sp\mrm{c}} \, ,
\qquad\quad
\langle w_{i_1}\cdots w_{i_n}\rangle_{\delta_\mrm{tot},\sp\mrm{c}}
=(-\im)^n\frac{\partial^n\ln Z}{\partial J_{i_1}\cdots\partial J_{i_n}}\bigg|_{\mbf{J}=0} \, ,
\ee
where repeated indices are summed over, and $i_1=1,2$, $i_2=1,2$, etc.
Here subscript $\delta_\mrm{tot}$ denotes the density-weighted
average~\eqref{eq:density-weighted-avg}, subscript `c' denotes
the connected part of the moment, and
$w_i=u_\|(\tau_i,\x_i)=\n_i\cdot\u(\tau_i,\x_i)$.
(Without loss of generality one may take the
lines of sight $\n_1$ and $\n_2$ to lie within the $xz$-plane,
as in Fig.~\ref{fig:config}.)
Then in terms of the connected moments
\be\label{eq:Z-exp}
Z(\J)
% =\ln\big\llangle\rme^{\im\k\cdot\Delta\u_\|}\big\rrangle
=\exp\bigg(\ssp\sum_{n=1}^\infty\frac{\im^n}{n!}\,
    J_{i_1}\cdots J_{i_n}\,
    \langle w_{i_1}\cdots w_{i_n}\rangle_{\delta_\mrm{tot},\sp\mrm{c}}\bigg) \, .
\ee
These expressions are general. As mentioned,
in the Gaussian streaming model we keep only the first and second 
connected moments, i.e.~truncating the sum at $n=2$. This leaves
the mean and covariance,
\bea
\bm\mu(\chi_1',\chi_2',\n_1\cdot\n_2)
\equiv
\langle\mbf{w}\rangle_{\delta_\mrm{tot},\sp\mrm{c}}
&=
\begin{pmatrix}
\langle u_\|(\chi_1')\rangle_{\delta_\mrm{tot},\sp\mrm{c}} \\[5pt]
\langle u_\|(\chi_2')\rangle_{\delta_\mrm{tot},\sp\mrm{c}}
\end{pmatrix} \, , 
\\[5pt]
\mbf{C}(\chi_1',\chi_2',\n_1\cdot\n_2)
\equiv
\langle\mbf{w}\mbf{w}^\T\rangle_{\delta_\mrm{tot},\sp\mrm{c}}
&=\begin{pmatrix}
\langle u_\|(\chi_1')\sp u_\|(\chi_1')\rangle_{\delta_\mrm{tot},\sp\mrm{c}} & \langle u_\|(\chi_1')\sp u_\|(\chi_2')\rangle_{\delta_\mrm{tot},\sp\mrm{c}} \\[5pt]
\langle u_\|(\chi_1')\sp u_\|(\chi_2')\rangle_{\delta_\mrm{tot},\sp\mrm{c}} & \langle u_\|(\chi_2')\sp u_\|(\chi_2')\rangle_{\delta_\mrm{tot},\sp\mrm{c}}
\end{pmatrix} \, .
\label{eq:C}
\eea
Here we have used the shorthand
$u_\|(\chi_i')\equiv u_\|(\tau_0-\chi_i',\chi_i'\n_i)$.
Note that the mean radial velocity $\bm\mu$, being density weighted, does not in general vanish.
Keeping terms in Eq.~\eqref{eq:Z-exp} up to second order
in $\J$ yields the generating function of a Gaussian:
\be\label{eq:ZG}
Z(\J)
=\exp\bigg(\im\sp\J\cdot\bm\mu-\frac12\J^\T\mbf{C}\sp\J \bigg) \, .
\ee
Inverse Fourier transform of $Z(\J)$, i.e.\ evaluating
Eq.~\eqref{eq:p-from-Z}, thus yields a two-dimensional Gaussian with
mean $\bm\mu$ and covariance $\mbf{C}$,
which when inserted back into Eq.~\eqref{eq:xis-los-int-bias}
furnishes the wide-angle Gaussian streaming model:%
\footnote{Since we are using spherical coordinates, the
probability distribution is perhaps better described as a
Maxwell--Boltzmann distribution $p\propto x^2\sp\rme^{-x^2}$
(or some two-dimensional analogue thereof).
Although for large variance we note that the Maxwellian is
well-approximated by a Gaussian (in the one-dimensional case).}
\be
\begin{split}
1+\xi_s(\chi_1,\chi_2,\n_1\cdot\n_2)
&=\frac{1}{\chi_1^2\sp\chi_2^2}
    \int^\infty_0\dif\chi_1'\:\chi_1'^{\sp2}
    \int^\infty_0\dif\chi'_2\:\chi_2'^{\sp2}\,
    \big[1+\xi_\mrm{tot}(\chi_1',\chi_2',\n_1\cdot\n_2)\big] \\
&\qquad\qquad
    \times\frac{1}{2\pi\sp|\mbf{C}|^{1/2}}\ssp
    % \times(2\pi)^{-1}|\mbf{C}(\r)|^{-1/2}\ssp
    \exp\bigg(\!
    -\frac12\big(\bm\chi-\bm\chi'-\bm\mu\big)^\T
    \mbf{C}^{-1}
    \big(\bm\chi-\bm\chi'-\bm\mu\big)
    \bigg) \, ,
    \label{eq:GSM-wide}
\end{split}
\ee
remembering that $\bm\mu$ and $\mbf{C}$ are functions of
$\chi_1'$, $\chi_2'$, and $\n_1\cdot\n_2$.
Note that this model does not assume that $\delta$ and $u_\|$ are Gaussian fields,
nor is it assuming that in the perturbative expansion~\eqref{eq:ZG}
the fields $\delta$ and $u_\|$ are small fluctuations.
Rather, this model is based on the \emph{correlations} being small on large scales. The Gaussian distribution arises from
our having truncated the generating function at second order in $\J=\bm\kappa$.
Of course, extensions to Eq.~\eqref{eq:GSM-wide} to include higher-order,
non-Gaussian statistics are also possible~\cite{Uhlemann:2015}.

The equivalent model without selection and evolution effects is obtained
by taking $\xi_\mrm{tot}\to\xi$ and $\delta_\mrm{tot}\to\delta$
in the density weighting. The above model also takes into account the
lookback time, which can be ignored by treating time in the usual way, i.e.\
as an independent variable (not degenerate with distance).
Overall, the effect of these three effects changes
the quantitative predictions but does not change the basic form of the model.

Equation~\eqref{eq:GSM-wide} is the full-sky generalisation of the well-known
Gaussian streaming model of the distant-observer limit:
\bea
1+\xi_s(s,\mu)
&=\int^\infty_{-\infty}\dif r_\|\,\big[1+\xi(r)\big]\cdot
    \frac{1}{\sqrt{2\pi}\sp\sigma_{12}(\r)}\,
    \exp\bigg(\!
    -\frac12\frac{(s_\|-r_\|-u_{12}(\r))^2}{\sigma^2_{12}(\r)}
    \bigg) \, ,
    \label{eq:GSM-DOL}
\eea
where $r_\|$ and $s_\|$ are the real- and redshift-space separations along
the line of sight; and
$u_{12}(\r)\equiv\langle\Delta u_\|\rangle_{\delta_\mrm{tot},\sp\mrm{c}}$ and
$\sigma^2_{12}(\r)\equiv\langle(\Delta u_\|)^2\rangle_{\delta_\mrm{tot},\sp\mrm{c}}$,
where $\Delta u_\|\equiv\n\cdot\u(\x_1)-\n\cdot\u(\x_2)$, are
the mean and dispersion of the pairwise velocity, respectively, and all
quantities are evaluated at a fixed time.
Although the wide-angle and distant-observer models are similar in form,
it requires some work to show that Eq.~\eqref{eq:GSM-wide} does indeed
reduce to Eq.~\eqref{eq:GSM-DOL} in the appropriate limit. We leave
the details of this calculation to Appendix~\ref{app:GSM-DOL}.

\section{Multipole expansion in the wide-angle regime}\label{sec:multipoles}
In this section we show that Eq.~\eqref{eq:xis-los-int}
recovers the standard linear predictions for the multipoles,
including those induced when going beyond the distant-observer limit.
Since the aim here is to compare our results with those in the wide-angle
literature, we will ignore contributions from galaxy evolution and
relativistic effects.

To facilitate the calculation, recall that the correlation function
on a fixed redshift slice can be expanded about the distant-observer
limit as~\cite{Reimberg_2016}
\be\label{eq:xi-s-multipole-exp}
\xi_s(s,\mu,d)
=\sum_{n=0}^\infty \Big(\frac{s}{d}\Big)^n
\sum_{\ell=0}^\infty
    \xi^{(n)}_\ell(s,d)\sp\mathcal{L}_\ell(\mu) \, ,
\ee
i.e.\ in terms of a small expansion parameter $\epsilon\equiv s/d$,
where for closely separated lines of sight a low-order expansion is valid.
Here $\mathcal{L}_\ell$ is Legendre polynomial of the $\ell$th degree,
$s=|\s_1-\s_2|$ is the separation, $\mu=\cos\theta$
(see Fig.~\ref{fig:config}), and $d$ is some distance to the
galaxy pair (to be made precise shortly).
Note that in addition to the explicit dependence of the multipoles in $d$ via $\epsilon$, $\xi_\ell^{(n)}$ depends also on $d$ through the evolution of the density and velocity, which depend on redshift, and therefore varies with distance $d$.
The usual Kaiser multipoles~\cite{Hamilton:1992zz} are given by the $n=0$ multipoles:
\begin{subequations}\label{eq:kaiser-multipoles}
\bea
\xi^{(0)}_0(s)
&=\int\frac{k^2\dif k}{2\pi^2}\,j_0(ks)
\bigg(\bA\bB P_{\delta\delta}(k)
-\frac{1}{3}(\bA+\bB)P_{\theta\delta}(k)
    +\frac{1}{5}P_{\theta\theta}(k)\bigg) \, , \\[2pt]
\xi^{(0)}_2(s)
&=\int\frac{k^2\dif k}{2\pi^2}\,j_2(ks)
\bigg(\frac{2}{3}(\bA+\bB)P_{\theta\delta}(k)
    -\frac{4}{7}P_{\theta\theta}(k)\bigg) \, , \\[2pt]
\xi^{(0)}_4(s)
&=\int\frac{k^2\dif k}{2\pi^2}\,j_4(ks)
    \frac{8}{35}P_{\theta\theta}(k) \, ,
\eea
\end{subequations}
where $\bA$ and $\bB$ are the linear galaxy bias of two tracers labelled A and B,
and $\theta=-f\delta$, where $f$ is the growth rate.
The wide-angle corrections are given by multipoles $n\geq1$, and
the $\ell$th multipole is given by the sum
$\xi_\ell(s,d)\equiv\sum_n \epsilon^n\sp\xi_\ell^{(n)}(s,d)$. 
Unlike in the distant-observer limit, the wide-angle contributions to the
multipoles depend on how the angular separation $\mu$ is defined, i.e.\
what we choose for the line of sight~\cite{Reimberg_2016,Castorina_2018}.

\subsection{Mid-point parametrisation}\label{sec:midpoint}
The shape and size of the multipoles depend on how we parametrise
the triangle as formed by the galaxy pair with the
observer. In particular, we need to fix the definition of
$\mu$. This means choosing a line of sight, and there is no
unique choice for this. In this work we use the line of sight
defined by the mid-point parametrisation
(see Fig.~\ref{fig:config}). This section describes this parametrisation
and collects some useful formulae. In Appendix~\ref{app:endpoint}
we give a formula relating multipoles in the mid-point parametrisation
to those in the end-point parametrisation.

First, the mid-point of the separation $\s=\s_1-\s_2$ is
given by $\mbf{d}\equiv(\s_1+\s_2)/2$.
We thus have $\s_1=\mbf{d}+\s/2$ and $\s_2=\mbf{d}-\s/2$. The
expansion parameter in Eq.~\eqref{eq:xi-s-multipole-exp} is
$\epsilon\equiv s/d$.
In particular, we may align $\mbf{d}$ with the $+z$-axis, i.e.\
$\hat{\mbf{d}}=\ez$.
We can also without loss of generality place the triangular
configuration in the $xz$-plane, with
the first galaxy placed in the left half-plane (with negative
$x$-coordinate) and the second galaxy placed in the
right half-plane (with positive $x$-coordinate); see Fig.~\ref{fig:config}.
With these choices $\hat\s=-\sqrt{1-\mu^2}\ssp\ex+\mu\ssp\ez$, with
$\mu\equiv\ez\cdot\hat\s$ and $\ex$, $\ez$ are unit vectors
along the $x$ and $z$ axes, respectively. Now
$\s_1=d\sp(\ez+\epsilon\sp\hat\s/2)$ and
$\s_2=d\sp(\ez-\epsilon\sp\hat\s/2)$, from which the unit vectors
are found to be
\begin{subequations}\label{eq:n1-and-n2}
\bea
\n_1&\equiv\hat\s_1=\Big(\ez+\frac{\epsilon}{2}\sp\hat\s\Big)
\sum_{n=0}^\infty %(-1)^\ell
\Big(-\frac{\epsilon}{2}\Big)^n
    \mathcal{L}_n(\mu)
=\ez-\frac\epsilon2\sqrt{1-\mu^2}\ssp\ex +\mathcal{O}(\epsilon^2)\, , \\
\qquad
\n_2&\equiv\hat\s_2=\Big(\ez-\frac{\epsilon}{2}\sp\hat\s\Big)
\sum_{n=0}^\infty \Big(+\frac{\epsilon}{2}\Big)^n
    \mathcal{L}_n(\mu)
=\ez+\frac\epsilon2\sqrt{1-\mu^2}\ssp\ex +\mathcal{O}(\epsilon^2) \, .
\eea
\end{subequations}
Observe that at $\mathcal{O}(\epsilon)$, the lines of sight
$\n_1$ and $\n_2$ are symmetric about the $z$-axis (equal and
opposite $x$-components).

Note the following relations when going between variables
$\{\chi_1,\chi_2,\cos\vartheta\equiv\n_1\cdot\n_2\}$ and $\{s,d,\mu\}$:
$s=(\chi^2_1+\chi^2_2-2\chi_1\chi_2\cos\vartheta)^{1/2}$,
$d=\frac12(\chi^2_1+\chi^2_2+2\chi_1\chi_2\cos\vartheta)^{1/2}$, and
$\mu^2
% =\frac14(\chi_1^2-\chi_2^2)^2/(|\s_1+\s_2||\s_1-\s_2|)^2
=\frac14(\chi_1^2-\chi_2^2)^2/\sp[(\chi_1^2+\chi_2^2)^2
-(2\chi_1\chi_2\cos\vartheta)^2]$. These follow from the cosine rule.

\subsection{Linear theory}\label{sec:recovery}
We now show that Eq.~\eqref{eq:xis-2-alt} recovers at zeroth-order
($n=0$) the standard Kaiser multipoles~\cite{Hamilton:1992zz}, 
and at the first-order ($n=1$) the wide-angle corrections.
As mentioned, the wide-angle corrections vanish at first-order in the
auto-correlation function but not for the cross-correlation function.
We will thus consider the cross-correlation between two tracer
populations, described by linear bias $\bA$ and $\bB$.
We will assume no magnification and evolution bias since we have already shown
in Section~\ref{sec:recover-lt} that we recover the correct linear expression
for $\delta_s$.
The details of our computations can be found in Appendix~\ref{app:recovery}.

First we convert Eq.~\eqref{eq:xis-2-alt} to the cross-correlation.
In linear theory this is done simply by
replacing $\delta(\x_1)\to\bA\sp\delta(\x_1)$ and $\delta(\x_2)\to\bB\sp\delta(\x_2)$:
\be%\label{eq:xis-exact}
1+\xi_s(s,\mu,d)
=\frac{1}{\chi_1^2\ssp\chi_2^2}
    \int\dif\chi_1'\,\chi_1'^{\sp2}\int\dif\chi_2'\,\chi_2'^{\sp2}
    \int\frac{\dif^2\bm\kappa}{(2\pi)^2}\,
    \rme^{-\im\bm\kappa\cdot(\bm\chi-\bm\chi')}
    \langle(1+\bA\sp\delta_1)(1+\bB\sp\delta_2)\ssp
    \rme^{\im\bm\kappa\cdot\mbf{w}}\rangle \, .
\ee
Here we have used the shorthands $\delta_1=\delta(\x_1)$ and
$\delta_2=\delta(\x_2)$, and as before
$\mbf{w}=(u_\|(\x_1),u_\|(\x_2))$. For convenience we will
also write
\be
U_i(\r)=\langle u_i(\x_1)\delta(\x_2)\rangle
\qquad\text{and}\qquad
\Psi_{ij}(\r)=\langle u_i(\x_1)u_j(\x_2)\rangle
\ee
for the velocity-density and velocity-velocity correlation functions.
Here the separation $\r=\x_1-\x_2$ is given in terms of 
the radial distances as $\r(\bm\chi')=\chi_1'\n_1-\chi_2'\n_2$,
and in redshift space $\s(\bm\chi)=\chi_1\n_1-\chi_2\n_2=\r(\bm\chi)$.

The idea of the calculation is to expand $\rme^{\im\bm\kappa\cdot\mbf{w}}$,
keeping up to quadratic terms and dropping zero-lag terms (which are absent
in the linear predictions). The integrations can then be done analytically
(see Appendix~\ref{app:recovery} for details). The result is
\be
\begin{split}
\xi_s(\chi_1,\chi_2,\n_1\cdot\n_2)
=\bA\bB\sp\xi(s)
   &-\bB\frac{\partial}{\partial\chi_1} U_i\sp\hat{n}_1^i
   +\bA\frac{\partial}{\partial\chi_2} U_i\sp\hat{n}_2^i
   +\frac{\partial}{\partial\chi_1}
    \frac{\partial}{\partial\chi_2}
    \Psi_{ij}\ssp\hat{n}_1^i\hat{n}_2^j \\[2pt]
&
   -\bB\frac{2}{\chi_1} U_i\sp\hat{n}_1^i
   +\bA\frac{2}{\chi_2} U_i\sp\hat{n}_2^i
   +\Big(\frac{2}{\chi_1}\frac{\partial}{\partial\chi_2}
    +\frac{2}{\chi_2}\frac{\partial}{\partial\chi_1}\Big)
        \Psi_{ij}\ssp\hat{n}_1^i\hat{n}_2^j \, .
    \label{eq:xis-Hk}
\end{split}
\ee
This is the linear correlation function corresponding to the right-hand
side of Eq.~\eqref{eq:dels-lin1}.
Here $\xi$, $U_i$ and $\Psi_{ij}$ depend on $\chi_1$ and $\chi_2$ through
$\s$ (and we remember that lines of sight are always constant with respect
to their radial derivatives, $\partial\sp\hat{n}^i/\partial\chi=0$).
The first line in Eq.~\eqref{eq:xis-Hk} yields the usual
Kaiser multipoles (among wide-angle corrections),
while the second line consists of terms suppressed by a factor of $\calH/k$
with respect to the Kaiser multipoles,
but are of the same order as the wide-angle contributions.

\subsubsection{Distant-observer limit}\label{sec:DOL-linear}
The multipoles of the distant-observer limit (i.e.~the Kaiser multipoles)
can be recovered by setting $\n_1=\n_2=\ez=(0,0,1)$ and taking
$\chi_1,\chi_2\to\infty$. Doing so eliminates the second line of terms in Eq.~\eqref{eq:xis-Hk}, leaving
\be
\xi_s(s,\mu)
=\bA\bB\sp\xi(s)
   -(\bA+\bB)\sp\partial_3 U_3(s,\mu)
   -\partial_3^2\Psi_{33}(s,\mu) \, ,
\qquad
   \text{(distant-observer limit)}
\ee
where derivatives are with respect to $s_3$, the $z$-component of $\s$.
This equation was first derived in Ref.~\cite{Fisher_1995}.
A straightforward computation of the derivatives yields
\begin{subequations}
\bea
\partial_3 U_3(s,\mu)
&=\int\frac{k^2\dif k}{2\pi^2}\,
    \bigg(\frac13\ssp j_0(ks) - \frac23\sp j_2(ks)\sp\mathcal{L}_2(\mu)\bigg)
    P_{\theta\delta}(k) \, , \label{eq:dU}\\
\partial_3^2\Psi_{33}(s,\mu)
&=-\int\frac{k^2\dif k}{2\pi^2}\,\bigg(\frac15\sp j_0(ks)
    -\frac47\sp j_2(ks)\mathcal{L}_2(\mu)
    +\frac{8}{35}\sp j_4(ks)\mathcal{L}_4(\mu)\bigg)P_{\theta\theta}(k) \, , \label{eq:d2Psi}
\eea
\end{subequations}
where $j_\ell$ is the $\ell$th-order spherical Bessel function.
From here it is not difficult to assemble the Kaiser multipoles~\eqref{eq:kaiser-multipoles}.

\subsubsection{Wide-angle corrections}\label{sec:wide-angle-corr}
The wide-angle corrections enter the $U_i$ terms at
order $\epsilon$ and the $\Psi_{ij}$ terms at $\epsilon^2$.
(Note that in the auto-correlation, i.e.\ when $\bA=\bB$, the
corrections also enter $U_i$ at $\epsilon^2$.)
Since we are interested only in the leading-order corrections
(order $\epsilon$), we need only focus on terms involving $U_i$ in
Eq.~\eqref{eq:xis-Hk}; the terms involving $\Psi_{ij}$ are as given in
the distant-observer limit so require no further calculation.
For the $\partial_i U_j$ terms in the first line of
Eq.~\eqref{eq:xis-Hk}, we have
\be
\begin{split}
-\bB\frac{\partial}{\partial\chi_1} U_i\sp\hat{n}_1^i
   +\bA\frac{\partial}{\partial\chi_2} U_i\sp\hat{n}_2^i
&=-(\bA+\bB)\int\frac{k^2\dif k}{2\pi^2}\,
    \bigg(\frac13\ssp j_0(ks) - \frac23\sp j_2(ks)\sp\mathcal{L}_2(\mu)\bigg)
    P_{\theta\delta}(k) \\[2pt]
&\qquad
    -\frac25\ssp\epsilon\ssp(\bA-\bB)\int\frac{k^2\dif k}{2\pi^2}\,
        \big(\mathcal{L}_1(\mu)-\mathcal{L}_3(\mu)\big)\sp
            j_2(ks) P_{\theta\delta}(k) \, ,
\end{split}
\ee
where the first integral on the right-hand side is the distant-observer
contribution, and the second integral is the leading-order wide-angle correction.\footnote{This agrees with equations (52) and (53) in 
Ref.~\cite{Bonvin:2013ogt}; see also equation (3.19) in Ref.~\cite{Bonvin:2018ckp}.}
The details of this computation can be found in Appendix~\ref{app:recovery}.
For the $U_i$ terms in Eq.~\eqref{eq:xis-Hk} up to leading
order in $\epsilon$, we have
\bea\label{eq:Uini}
   -\bB\frac{2}{\chi_1} U_i\sp\hat{n}_1^i
   +\bA\frac{2}{\chi_2} U_i\sp\hat{n}_2^i
&=2\sp\epsilon\sp(\bA-\bB)
    \int\frac{k^2\dif k}{2\pi^2}
    \frac{j_1(ks)}{ks}\mathcal{L}_1(\mu) P_{\theta\delta}(k) \, .
\eea
The leading-order wide-angle multipoles are thus
\begin{subequations}\label{eq:dip-oct}
\bea
\xi^{(1)}_1(s)
&=(\bA-\bB)\int\frac{k^2\dif k}{2\pi^2}\,
    \bigg(\!-\frac25\sp j_2(ks)
        + 2\ssp\frac{j_1(ks)}{ks}\bigg) P_{\theta\delta}(k) \, , \\[2pt]
\xi^{(1)}_3(s)
&=\frac25(\bA-\bB)\int\frac{k^2\dif k}{2\pi^2}\,
    j_2(ks)P_{\theta\delta}(k) \, .
\eea
\end{subequations}
These are consistent with those given in, e.g.\
Refs.~\cite{Bonvin:2018ckp,Taruya_2019}.
Note that when working within the end-point parametrisation the odd multipoles, 
$\xi_1^{(1)}$ and $\xi_3^{(1)}$, receive additional contributions, which are of
a geometric, non-cosmological nature (Appendix~\ref{app:endpoint}).
Thus we have recovered the linear multipoles.

\section{Conclusions}\label{sec:conclusions}
We have described a framework to model in the nonlinear regime
not only wide-angle effects but also selection and
relativistic effects.
Our main result is Eq.~\eqref{eq:xis-los-int-bias}, an expression for
the redshift-space correlation function which
is valid in both the nonlinear regime and on the full sky, and
accounts for the survey flux limit and the population evolution of tracers.
Based on this expression, we have also given the full-sky generalisation
of the Gaussian streaming model, Eq.~\eqref{eq:GSM-wide}, which we
have checked reduces to the well-known
flat-sky model~\eqref{eq:GSM-DOL} in the appropriate limit.

The correlation function~\eqref{eq:xis-los-int-bias}
takes a lensing-like form (i.e.\ is given by integrals along each
line of sight) which
can be understood probabilistically: a given two-point correlation 
in redshift space is determined by averaging over all the possible
two-point correlations in real space that can be formed on the two 
lines of sight. Geometrically, this can be understood as a weighted sum over
the space of triangular configurations in which the observer is fixed at
one vertex with the galaxies at the other two (at the ends of the lines of sight).
Since the opening angle is fixed, the probability space is two dimensional
and given by the joint statistics of the line-of-sight components of
the galaxy velocities.
We note that this heuristic generalises to higher-order correlation functions
(e.g.\ for the three-point function the sum is over tetrahedrons).

We have also given a non-perturbative expression~\eqref{eq:dels-los-int}
for the overdensity in redshift space.
Performing a perturbative expansion of this expression, we showed that
we are able to recover all but two terms of the well-known
linear expression of the overdensity at subleading order;
see Eq.~\eqref{eq:dels-lin-all}.
The first term missing traces back to the
fact that observations probe the number density of galaxies not in their
rest frame but in a frame tilted with respect to it. This results in an
additional kinematic term but requires a covariant expression of number 
conservation.
The second term is the gravitational redshift, whose absence is due
to the simple fact that we have chosen to exhibit the formalism using
the familiar redshift mapping~\eqref{eq:sx}.
A model of the overdensity, complete down to $\mathcal{O}(\calH/k)$
effects, will be presented in a follow-up work~\cite{Dam:2023}.
Nevertheless, the expression we have derived provides a compact
description of a large number of terms (RSD, magnification bias,
evolution bias, projection effects related to the lightcone, etc). 
Furthermore, our work provides a simple quasi-Newtonian derivation
to the full relativistic calculation.

In summary, we have shown that the streaming model is not limited
to the distant-observer limit but that it can be straightforwardly extended
into the wide-angle regime and be built upon to include a number of other
important effects.
In a future work we will present numerical results for a realistic model
including nonlinear evolution and galaxy bias, with a view towards an
eventual measurement of the gravitational redshift.

\subsection*{Acknowledgements}

This work is supported by the European Research Council (ERC) under the European Union’s Horizon 2020 research and innovation program (Grant agreement No.~863929; project title ``Testing the law of gravity with novel large-scale structure observables". CB acknowledges support from the Swiss National Science Foundation.

% \clearpage
\appendix

\section{On the mean density in the wide-angle regime}\label{app:ns}
In Section~\ref{sec:wide} we assumed that
$\bar{n}_s\equiv\langle n_s(\s)\rangle=\bar{n}$.
However, this is not assured in the wide-angle regime.
This can be shown by direct calculation of the expectation
of Eq.~\eqref{eq:ns}:
\be
\langle n_s(\s)\rangle
=\frac{1}{\chi^2}\ssp\bar{n}
    \int^\infty_0\dif\chi'\sp\chi'^{\sp2}
    \int\frac{\dif k}{2\pi}\,\rme^{-\im k(\chi-\chi')}
    \big\langle[1+\delta(\chi'\n)]\ssp\rme^{\im k u_\|}\big\rangle \, .
\ee
In general, the right-hand side does \emph{not} evaluate to $\bar{n}$.
We can see this as follows.
By the cumulant expansion formula
$\langle\rme^{iX}\rangle=\exp\langle\rme^{iX}\rangle_\mrm{c}$, we have that
$\langle\rme^{\im k\sp u_\|}\rangle=\rme^{-k^2\sigma_u^2/2}$
and $\langle\rme^{\im k\sp u_\|}\delta\rangle=0$,
assuming as a first approximation that $\delta$ and $u_\|$ are Gaussian fields.
Recognising that $\rme^{-k^2\sigma_u^2/2}$ is the Fourier transform of a 
Gaussian with mean zero and variance $\sigma_u^2$, we have
\be\label{eq:nbar_s-avg}
\langle n_s(\s)\rangle
=\frac{1}{\chi^2}\,\bar{n}
    \int^\infty_0\dif\chi'\sp\chi'^{\sp2}
    \frac{1}{\sqrt{2\pi}\sp\sigma_u}\,
    \rme^{-(\chi-\chi')^2/2\sp\sigma_u^2} \, .
\ee
That the mean density $\langle n_s(\s)\rangle$ is a position-dependent
quantity may seem strange at first, but this is just a consequence of
the loss of statistical homogeneity in the wide-angle regime, with
the observer representing a preferred location in space.
Indeed, in the distant-observer limit, where
homogeneity is retained, this $\chi$ dependence drops out:
if $\chi\gg\sigma_u$, then the Gaussian in the 
integrand is sharply peaked around $\chi'=\chi$, so that the
integral evaluates to approximately $\chi^2$.
Therefore $\langle n_s\rangle\to\bar{n}$ as $\chi/\sigma_u\to\infty$,
so that it is perfectly
valid to take $\langle n_s\rangle=\bar{n}$ in this limit.
But short of this limit there are corrections, which decrease with
depth. Fortunately, convergence to this limit is rapid.
Quantitatively, with the $\Lambda$CDM value
$\sigma_u=5.8\ssp h^{-1}\mrm{Mpc}$ (corresponding to a velocity dispersion
of about $300\ssp\mrm{km\,s^{-1}}$), we find for depths
$\chi\geq100\ssp h^{-1}\mrm{Mpc}$ (or $z\geq0.023$) that the
deviations from $\langle n_s(\s)\rangle/\bar{n}=1$ are $\leq0.3\%$,
i.e.\ small in most situations of interest.

There are of course corrections to these estimates from non-Gaussianity due to nonlinear gravitational evolution.
These corrections are largest on small scales $k\sim1/\chi$. On
intermediate scales $k\sim0.1\,h\,\mrm{Mpc}^{-1}$,
where nonlinearities begin to be important, we expect perturbatively
small departures from Gaussianity.
This translates to non-Gaussianities becoming important
at depths $\chi\simeq60\ssp h^{-1}\mrm{Mpc}$ ($z\sim0.01$) or
shallower, i.e.\ small or negligible by the time we reach the
convergence scale $\chi\simeq100\ssp h^{-1}\mrm{Mpc}$. This means
that, unless one's sample contains very local galaxies, there seems
little harm in taking $\langle n_s(\s)\rangle=\bar{n}$
(though one can always include the corrections should they be
wanted).

\section{Recovering the standard Gaussian streaming model of the distant-observer limit}
\label{app:GSM-DOL}
In this appendix we verify that the usual distant-observer
streaming model~\eqref{eq:GSM-DOL} is recovered as a special case of
the full-sky streaming model~\eqref{eq:GSM-wide}. Clearly we
must end up with one less integral, leaving an integral over
the separation $r_\|$.
However, this is not as straightforward as simply taking the distant-observer
limit, $\chi_1,\chi_2\to\infty$ and $\n_1\to\n_2$. 
It turns out to be convenient to centre the coordinates on the
redshift-space positions, with the coordinate transformation
$\chi_1'=\chi_1-q_1$
and $\chi_2'=\chi_2-q_2$, or $\bm\chi'=\bm\chi-\mbf{q}$.
Equation~\eqref{eq:GSM-wide} then reads
\be
\begin{split}
1+\xi_s(\chi_1,\chi_2,\n_1\cdot\n_2)
&=\int^{\chi_1}_{-\infty}\dif q_1\,
    \Big(1-\frac{q_1}{\chi_1}\Big)^2
    \int^{\chi_2}_{-\infty}\dif q_2\,
    \Big(1-\frac{q_2}{\chi_2}\Big)^2
    [1+\xi(r)] \\
&\quad
    \times\frac{1}{2\pi|\mbf{C}(\r)|^{1/2}}\ssp
    \exp\bigg(\!
    -\frac12\big(\q-\bm\mu(\r)\big)^\T
    \mbf{C}^{-1}(\r)
    \big(\q-\bm\mu(\r)\big)
    \bigg) \, .
\end{split}
\ee
(Here we have ignored for simplicity the selection effects and the lookback time
so that $\bm\mu$ and $\mbf{C}$ depend on a triangle configuration parametrised
by $\r$, i.e.\ we are working on a constant-time hypersurface.)
In the limit $\chi_1,\chi_2\to\infty$ the first two factors in parentheses
tend to unity (noting that at large $q_1,q_2$ these factors become
irrelevant as the Gaussian rapidly takes the whole integrand to zero). Thus, setting these
factors to unity, and doing some straightforward matrix algebra, the foregoing 
expression becomes
\be\label{eq:xis-2to1}
1+\xi_s
=\int^\infty_{-\infty}\dif q_1
    \int^\infty_{-\infty}\dif q_2\,
    \frac{1+\xi(r)}{2\pi\sigma^2\sqrt{1-\rho^2}}\,
    \exp\bigg(\!
    -\frac12\frac{(\Delta_1-\Delta_2)^2+2(1-\rho)\Delta_1\Delta_2}{\sigma^2(1-\rho^2)}
    \bigg) \, ,
\ee
where as shorthands $\Delta_1\equiv q_1-\mu_1$ and $\Delta_2\equiv q_2-\mu_2$,
while $C_{11}=C_{22}=\sigma^2$ and $C_{12}=\rho\sp\sigma^2$,
where $\sigma=\sigma(\r)$ and $\rho=\rho(\r)$ (or functions of $q_1$ and $q_2$).
Defining the line-of-sight separation in real and redshift space,
$r_\|=\chi_1'-\chi_2'$ and $s_\|=\chi_1-\chi_2$, and the mid-points
$\bar{r}_\|=(\chi_1'+\chi_2')/2$ and $\bar{s}_\|=(\chi_1+\chi_2)/2$,
we have $q_1-q_2=s_\|-r_\|$ and $(q_1+q_2)/2=\bar{s}_\|-\bar{r}_\|$, which implies
$q_1=\bar{s}_\|-\bar{r}_\|+\tfrac12(s_\|-r_\|)$ and
$q_2=\bar{s}_\|-\bar{r}_\|-\tfrac12(s_\|-r_\|)$.
(Note that $s_\|$ and $\bar{s}_\|$ are fixed by the
redshift-space configuration.)
Making another change of coordinates, $(q_1,q_2)$ to $(r_\|,\bar{r}_\|)$,
we have after some more algebra
\be
1+\xi_s(s_\|,s_\perp)
=\int^\infty_{-\infty}\dif r_\|\,
    \frac{1+\xi(r)}{2\pi\sigma^2\sqrt{1-\rho^2}}\,
    \exp\bigg(\!
    -\frac12\frac{(s_\|-r_\|-u_{12})^2}{2\sp\sigma^2(1-\rho)}\bigg)
    \int^\infty_{-\infty}\dif \bar{r}_\|\,
    \exp\bigg(\!
    -\frac{(\bar{s}_\|-\bar{r}_\|)^2}{\sigma^2(1+\rho)}
    \bigg) \, ,
\ee
where we recognised that
$\mu_1-\mu_2
=\langle u_\|(\x_1)\rangle_{\delta,\mrm{c}}-\langle u_\|(\x_2)\rangle_{\delta,\mrm{c}}\equiv u_{12}$,
and noted that 
$u_{12}$, $\sigma$, and $\rho$ depend on $r_\|$, but not $\bar{r}_\|$, hence
the last integral.
Here we have $r^2=r_\|^2+r_\perp^2$ and $r_\perp=s_\perp$.
Upon doing the last (Gaussian) integral over $\bar{r}_\|$ and noting that
$2\sp\sigma^2(1-\rho)=2C_{11}-2C_{12}=\langle(\Delta u_\|)^2\rangle_{\delta,\mrm{c}}
\equiv\sigma_{12}^2$, we hence recover Eq.~\eqref{eq:GSM-DOL}, the usual
Gaussian streaming model, i.e.~in the distant-observer limit.

\section{Linear theory multipoles}\label{app:recovery}
In this appendix we calculate the contributions
to the multipoles from wide-angle effects and inverse-distance terms
(from the selection function), filling in some of the details of
Section~\ref{sec:recovery}.
We will compute from Eq.~\eqref{eq:xis-los-int} the leading-order
wide-angle corrections, i.e.~at
$\mathcal{O}(\epsilon)$, and will consider the cross-correlation of
two different tracers, described by linear bias
$\bA$ and $\bB$. For this calculation it is convenient to start with 
Eq.~\eqref{eq:xis-2-alt}, which for the cross-correlation function is given by simply 
replacing $\delta(\x_1)\to\bA\sp\delta(\x_1)$ and $\delta(\x_2)\to\bB\sp\delta(\x_2)$:
\be\label{eq:xis-exact}
1+\xi_s(s,\mu,d)
=\frac{1}{\chi_1^2\chi_2^2}
    \int\chi_1'^{\sp2}\dif\chi_1'\int\chi_2'^{\sp2}\dif\chi_2'
    \int\frac{\dif^2\bm\kappa}{(2\pi)^2}\,
    \rme^{-\im\bm\kappa\cdot(\bm\chi-\bm\chi')}
    \langle(1+\bA\sp\delta_1)(1+\bB\sp\delta_2)\sp
    \rme^{\im\bm\kappa\cdot\mbf{w}}\rangle \, .
\ee
Here we have used the shorthands $\delta_1=\delta(\x_1)$ and
$\delta_2=\delta(\x_2)$, and as before
$\mbf{w}=(\n_1\cdot\u(\x_1),\n_2\cdot\u(\x_2))$. For convenience we write
$U_i(\r)\equiv\langle u_i(\x_1)\delta(\x_2)\rangle$ and
$\Psi_{ij}(\r)=\langle u_i(\x_1)u_j(\x_2)\rangle$
for the velocity--density and velocity--velocity correlation functions.
In terms of the (linear) power spectra,
\bea
U_i(\s)
&=\int\frac{\dif^3\k}{(2\pi)^3}\,\rme^{-\im\k\cdot\s}\ssp
\frac{\im k_i}{k^2}\ssp P_{\theta\delta}(k)
=-\partial_i\int\frac{k^2\dif k}{2\pi^2}\,\frac{1}{k^2}\, j_0(ks)
    P_{\theta\delta}(k) \, , \label{eq:Ui} \\
\Psi_{ij}(\s)
&=\int\frac{\dif^3\k}{(2\pi)^3}\,\rme^{-\im\k\cdot\s}\ssp
    \frac{\im k_i}{k^2}\frac{-\im k_j}{k^2}\ssp P_{\theta\theta}(k) 
=-\partial_i\partial_j
    \int\frac{k^2\dif k}{2\pi^2}\frac{1}{k^4}\,j_0(ks)
    P_{\theta\theta}(k) \, ,
\eea
where $\theta$ is the velocity divergence,
$\partial_i=\partial/\partial s_i$, and we have used that
$u_i=\im k_i/k^2\theta$ (for a potential flow).
Recall that the separation $\r=\x_1-\x_2$ is given in terms of 
the radial distances as $\r(\bm\chi')=\chi_1'\n_1-\chi_2'\n_2$,
and in redshift space $\s(\bm\chi)=\chi_1\n_1-\chi_2\n_2=\r(\bm\chi)$.

To evaluate Eq.~\eqref{eq:xis-exact}, we expand the generator and
keep only up to quadratic terms:
\bea
% [1+\xi(r)]\llangle\rme^{\im\bm\kappa\cdot\bm\nu}\rrangle
\langle(1+\bA\delta_1)(1+\bB\delta_2)\sp\rme^{\im\bm\kappa\cdot\mbf{w}}\rangle
&\simeq1+\xi(r)+\im\kappa_a\langle w_a(\bA\delta_1+\bB\delta_2)\rangle
    -\tfrac12\sp\kappa_a\kappa_b\langle w_a w_b\rangle
    \nonumber\\
&=1+\xi(r)+\im(\bB\kappa_1\sp\hat{n}_1^i-\bA\kappa_2\sp\hat{n}_2^i)\sp U_i(\r)
    -\kappa_1\sp\kappa_2\sp\hat{n}_1^i\sp\hat{n}_2^j\sp\Psi_{ij}(\r) \, ,
    \label{eq:gen-exp2}
\eea
where we have dropped zero-lag terms since they are absent in the linear
predictions. Here we have used that
$\langle\u_2\delta_1\rangle=-\langle\u_1\delta_2\rangle$, and
$\langle\u_1\delta_1\rangle=\langle\u_2\delta_2\rangle=0$, which
follow from isotropy of the underlying fields.

We will now evaluate Eq.~\eqref{eq:xis-exact} using the
expansion~\eqref{eq:gen-exp2}. This is a two-step calculation:
first evaluate the $\bm\kappa$ integral to yield a Dirac delta function,
then evaluate the radial integrals.
First, focus on the $U_i$ term in Eq.~\eqref{eq:gen-exp2};
applying the $\kappa_i$ integral on this we have
\be\nonumber
\int\frac{\dif^2\bm\kappa}{(2\pi)^2}\,
    \rme^{-\im\bm\kappa\cdot(\bm\chi-\bm\chi')}\,
    \im(\bB\kappa_1\n_1-\bA\kappa_2\n_2)\cdot\U(\r)
=\U(\r)\cdot\Big(\bA\sp\n_2\frac{\partial}{\partial\chi_2}
    -\bB\sp\n_1\frac{\partial}{\partial\chi_1}\Big)
    \ssp\delD(\bm\chi-\bm\chi') \, .
\ee
% where we recall that $\r$ depends on $\chi_1'$ and $\chi_2'$
% (not $\chi_1$ and $\chi_2$).
Inserting this back into the line-of-sight integrals and doing the integration
with the help of the delta functions, we obtain
\bea\nonumber
\frac{1}{\chi_1^2\sp\chi_2^2}
\Big(\bA\frac{\partial}{\partial\chi_2}\n_2
    -\bB\frac{\partial}{\partial\chi_1}\n_1\Big)
    \cdot\chi_1^{2}\ssp\chi_2^{2}\ssp
    \U(\s)
&=\bA\frac{1}{\chi_2^2}
\frac{\partial}{\partial\chi_2}(\chi_2^2\sp\U)\cdot\n_2
    -\bB\frac{1}{\chi_1^2}\frac{\partial}{\partial\chi_1}(\chi_1^2\sp\U)\cdot\n_1 \, .
% =\bm\nabla\cdot(\mbf{R}\U)\big|^2_1
\eea
Note that $\partial\sp s/\chi_1=\n_1\cdot\bm\nabla s=\n_1\cdot\hat\s$
and $\partial\sp s/\chi_2=-\n_2\cdot\bm\nabla s=-\n_2\cdot\hat\s$.
Next, the $\Psi_{ij}$ term in Eq.~\eqref{eq:gen-exp2};
plugging this into the $\kappa_i$ integral gives
\bea\nonumber
-\frac12\int\frac{\dif^2\bm\kappa}{(2\pi)^2}\,
    \rme^{-\im\bm\kappa\cdot(\bm\chi-\bm\chi')}\,
    \kappa_1\sp\kappa_2\sp\hat{n}_1^i\sp\hat{n}_2^j\sp\Psi_{ij}(\r)
&=\Psi_{ij}(\r)\sp\hat{n}_1^i\hat{n}_2^j\ssp
    \frac{\partial}{\partial\chi_1}\frac{\partial}{\partial\chi_2}
        \delD(\bm\chi-\bm\chi') \, .
\eea
Inserting this back into Eq.~\eqref{eq:xis-exact} and doing the
radial integrals we have
\be\nonumber
% \frac{1}{\chi_1^2\chi_2^2}\frac{\partial}{\partial\chi_1}\frac{\partial}{\partial\chi_2}\chi_1^2\chi_2^2\sp
%     \hat{n}_1^i\sp\hat{n}_2^j\sp
%     \Psi_{ij}(\s)
\frac{1}{\chi_1^2}\frac{\partial}{\partial\chi_1}\chi_1^2\,
    \frac{1}{\chi_2^2}\frac{\partial}{\partial\chi_2}\chi_2^2\,
    \Psi_{ij}(\s)\sp\hat{n}_1^i\hat{n}_2^j
=
\bigg(\frac{\partial}{\partial\chi_1}\frac{\partial}{\partial\chi_2}
    +\frac{2}{\chi_1}\frac{\partial}{\partial\chi_2}
    +\frac{2}{\chi_2}\frac{\partial}{\partial\chi_1}
    +\frac{2}{\chi_1}\frac{2}{\chi_2}
    \bigg)\Psi_{ij}(\s)\sp\hat{n}_1^i\hat{n}_2^j \, ,
\ee
where the second and third term on the right-hand side are order $\calH/k$,
while the last is order $(\calH/k)^2$.

Altogether we have
\be\label{eq:xis-spherical}
\xi_s(\chi_1,\chi_2,\n_1\cdot\n_2)
=\bA\bB\sp\xi(s)
   -\bB\frac{1}{\chi_1^2}\frac{\partial}{\partial\chi_1}
    \chi_1^2\sp U_i\sp\hat{n}_1^i
   +\bA\frac{1}{\chi_2^2}\frac{\partial}{\partial\chi_2}
    \chi_2^2\sp U_i\sp\hat{n}_2^i
   +\frac{1}{\chi_1^2}\frac{\partial}{\partial\chi_1}\chi_1^2\,
    \frac{1}{\chi_2^2}\frac{\partial}{\partial\chi_2}\chi_2^2\,
    \Psi_{ij}\ssp\hat{n}_1^i\hat{n}_2^j \, ,
\ee
where $\xi$, $U_i$ and $\Psi_{ij}$ depend on $\chi_1$ and $\chi_2$ through
$\s$, and we remember that
$\partial\sp\hat{n}^i/\partial\chi=0$, i.e.~lines of sight are
always constant with respect to their radial derivatives.
This equation is the wide-angle formula for the linear correlation
function and as we just saw the last derivative produces an
order $(\calH/k)^2$ term that we will henceforth ignore.
Evaluating the derivatives in Eq.~\eqref{eq:xis-spherical}
yields Eq.~\eqref{eq:xis-Hk} in the main text.

We now move onto computing the wide-angle corrections. For this
it is convenient to switch to Cartesian coordinates, noting
that for any function $f(\s)$, with $\s=\s_1-\s_2$,
we have by the chain rule
% $\partial f/\partial\sp\s_1=\partial f/\partial\sp\s$
% and $\partial f/\partial\sp\s_2=-\partial f/\partial\sp\s$.
% Moreover, 
$\partial f/\partial\chi_1=\hat{n}_1^j\sp\partial_j f$
and $\partial f/\partial\chi_2=-\hat{n}_2^j\sp\partial_j f$,
since $\partial s_1^j/\partial\chi_1=\hat{n}_1^j$
and $\partial s_2^j/\partial\chi_2=\hat{n}_2^j$. With these,
Eq.~\eqref{eq:xis-Hk} becomes
\be
\begin{split}
\xi_s(\chi_1,\chi_2,\n_1\cdot\n_2)
=\bA\bB\sp\xi(s)
   &-\big(\bB\sp\hat{n}_1^i\sp\hat{n}_1^j+\bA\sp\hat{n}_2^i\sp\hat{n}_2^j\sp\big)\sp
    \partial_i U_j
   -\hat{n}_1^i\sp\hat{n}_2^j\sp\hat{n}_1^k\sp\hat{n}_2^l\ssp
    \partial_k\partial_l\Psi_{ij}
    \\[3pt]
&
   -\bB\frac{2}{\chi_1} U_i\sp\hat{n}_1^i
   +\bA\frac{2}{\chi_2} U_i\sp\hat{n}_2^i
   +\Big(\frac{2}{\chi_1}\sp\hat{n}_1^k
    -\frac{2}{\chi_2}\sp\hat{n}_2^k\Big)\ssp
    \partial_k\Psi_{ij}\ssp
        \hat{n}_1^i\hat{n}_2^j \, ,
    \label{eq:xis-Hk2}
\end{split}
\ee
where as a shorthand $\partial_i=\partial/\partial s_i$.
Note that when $\bA=\bB$ the wide-angle corrections enter terms in the
first line at second order in $\epsilon$, and when $\bA\neq\bB$ they enter
at first order in $\epsilon$.

It now remains to compute the multipoles. We will first compute the zeroth-order
multipoles, i.e.~the usual Kaiser multipoles, then the first-order
multipoles that are associated with the wide-angle contributions.
The following derivatives will be needed:
\begin{subequations}\label{eq:j0-diff}
\bea
\frac{1}{k^2}\,\partial_m\partial_n\, j_0(ks)
&=-\frac{j_1(ks)}{ks}\delta_{mn} + j_2(ks)\ssp\hat{s}_m\hat{s}_n \, , 
\label{eq:D2j0} \\[1pt]
\frac{1}{k^3}\,\partial_j\partial_m\partial_n\, j_0(ks)
&=\frac{j_2(ks)}{ks}(\hat{s}_j\delta_{mn}+\text{2 perm.})
    -j_3(ks)\ssp\hat{s}_j\hat{s}_m\hat{s}_n \, , \label{eq:D3j0} \\[1pt]
\frac{1}{k^{4}}\,{\partial_i \partial_j \partial_m \partial_n} \, j_0(ks)
&=\frac{j_2(ks)}{(ks)^2}(\delta_{ij}\delta_{mn} + \text{2 perm.}) 
    - \frac{j_3(ks)}{ks}(\hat{s}_i\hat{s}_j\delta_{mn} + \text{5 perm.})
    + j_4(ks)\ssp\hat{s}_i\hat{s}_j\hat{s}_m\hat{s}_n \, ,
    \label{eq:D4j0}
\eea
\end{subequations}
where $\partial_i\equiv\partial/\partial s_i$.
Note that $(2\ell+1)\sp j_\ell(x)/x=j_{\ell-1}(x)+j_{\ell+1}(x)$.

\subsection{Distant-observer limit}
To recover the Kaiser multipoles, set $\n_1=\n_2=\ez=(0,0,1)$.
Then Eq.~\eqref{eq:xis-Hk2} simplifies to
\be\label{eq:xis-DOL}
\xi_s(\chi_1,\chi_2,\n_1\cdot\n_2)
=\bA\bB\sp\xi(s)
   -(\bA+\bB)\sp\partial_3 U_3
   -\partial_3^2\Psi_{33}
   -\Big(\bB\frac{2}{\chi_1}-\bA\frac{2}{\chi_2}\Big)\sp U_3
   +\Big(\frac{2}{\chi_1}-\frac{2}{\chi_2}\Big)\sp\partial_3\Psi_{33} \, .
\ee
In the distant-observer limit, we can immediately discard all terms order
$\epsilon$ and higher, namely, the last two terms
in Eq.~\eqref{eq:xis-DOL}, since with $\chi\sim d$ and $U\sim\partial\Psi\sim s$
they are $\mathcal{O}(\epsilon)$.
The remaining terms evaluate to Eqs.~\eqref{eq:dU} and \eqref{eq:d2Psi}
in the main text,
and from these equations it is straightforward to assemble the Kaiser 
multipoles~\eqref{eq:kaiser-multipoles}.
Note that for the auto-correlation function ($\bA=\bB$),
wide-angle effects enter the multipoles at $\epsilon^2$, not
order $\epsilon$.
This is only true in the mid-point (and bisector) parametrisations, however.

\subsection{Wide-angle contributions at leading order}
Referring back to Eq.~\eqref{eq:xis-Hk2}, wide-angle corrections enter the
terms $U$ and $\partial U$ at order $\epsilon$.
By contrast, wide-angle corrections enter the $\partial\Psi$ and $\partial^2\Psi$ terms at
order $\epsilon^2$, so do not need to be considered further in this
leading-order calculation.

The fact that the corrections are not all second order is a consequence of the bias
parameters spoiling invariance of the correlations under pair interchange.
We thus need only focus on the $U_i$ terms.
To develop Eq.~\eqref{eq:xis-Hk2} further we use the leading-order
expressions $\n_1=\ez+\frac\epsilon2\sp\mbf{n}^{(1)}$ and
$\n_2=\ez-\frac\epsilon2\sp\mbf{n}^{(1)}$, where
$\mbf{n}^{(1)}=-\sqrt{1-\mu^2}\sp\ex$. For the $\partial_i U_j$ term in 
Eq.~\eqref{eq:xis-Hk2} we have, with the help of Eq.~\eqref{eq:D2j0},
\be
\begin{split}
\big(\bB\sp\hat{n}_1^i\sp\hat{n}_1^j+\bA\sp\hat{n}_2^i\sp\hat{n}_2^j\sp\big)\sp
    \partial_i U_j
% &=(\bA+\bB)R^{(0)}_{33}\partial_3 U_3
% +\frac\epsilon2(\bB-\bA)
%     \big(R_{13}^{(1)}\partial_1U_3+R_{13}^{(1)}\partial_3 U_1\big)
%     \nonumber\\
&=(\bA+\bB)\int\frac{k^2\dif k}{2\pi^2}\,
    \bigg(\frac13\ssp j_0(ks) - \frac23\sp j_2(ks)\sp\mathcal{L}_2(\mu)\bigg)
    P_{\theta\delta}(k) \\
&\qquad
    +\frac25\ssp\epsilon\ssp(\bB-\bA)\int\frac{k^2\dif k}{2\pi^2}\,
        \big(-\mathcal{L}_1(\mu)+\mathcal{L}_3(\mu)\big)\sp
            j_2(ks) P_{\theta\delta}(k) \, ,
\end{split}
\ee
where the second term in this expression is the wide-angle correction
(which agrees with equations 52 and 53 in Ref.~\cite{Bonvin:2013ogt};
see also equation 3.19 in Ref.~\cite{Bonvin:2018ckp}).
For the $U_i$ terms in Eq.~\eqref{eq:xis-Hk} we use Eq.~\eqref{eq:Ui} and
contract with the appropriate line of sight. The result is Eq.~\eqref{eq:Uini}.
Gathering these results together, it is straightforward exercise to
construct the multipoles~\eqref{eq:dip-oct}.

\section{End-point parametrisation}\label{app:endpoint}
The end-point parametrisation is less symmetric than the mid-point
parametrisation (it induces odd multipoles) but is
often preferred for practical reasons, e.g.~for power-spectrum  
estimation~\cite{Scoccimarro:2015,Beutler2018:1810.05051}.
For completeness, in this appendix we derive the relation between
the multipoles in the mid-point parametrisation (used in this work)
and that in the end-point parametrisation, denoted
$\xi_\ell$ and $\xi_\ell^\mrm{ep}$, respectively. 

\begin{figure}[!t]
  \centering
  \includegraphics[scale=1]{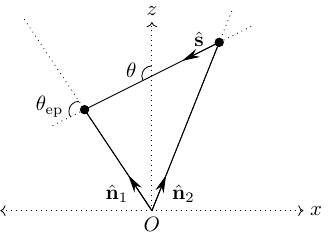}
  \caption{Comparison of end-point separation
    $\cos\theta_\mrm{ep}=\mu_\mrm{ep}=\n_1\cdot\hat\s$ and
    the mid-point separation $\cos\theta=\mu=\ez\cdot\hat\s$.}
\label{fig:config-EP}
\end{figure}

In general, the (cosine of the) angular separation can be defined as 
$\mu=\hat{\mbf{d}}\cdot\hat\s$. In the mid-point parametrisation
$\hat{\mbf{d}}=\ez$, while in the end-point parametrisation $\hat{\mbf{d}}=\n_1$
(or alternatively $\hat{\mbf{d}}=\n_2$); see Fig.~\ref{fig:config-EP}.
Based on a trigonometric analysis of Fig.~\ref{fig:config},
we find that the separations are related by 
% \be% \mu'=\mu+\Big(\frac{\epsilon}{2}\Big)(\mu^2-1)+\mathcal{O}(\epsilon^2),
% \ee
\be
\mu=\mu_\mrm{ep}+\frac\epsilon2\ssp(\mu'^2-1)+\mathcal{O}(\epsilon^2)\, ,
\qquad
\mu_\mrm{ep}=\mu-\frac\epsilon2\ssp(\mu^2-1)+\mathcal{O}(\epsilon^2) \, .
\ee
The expansion parameter in the end-point parametrisation is
$\epsilon_\mrm{ep}\equiv s/s_1$ and
$\epsilon_\mrm{ep}=\epsilon+\mathcal{O}(\epsilon^2)$; since we will
be working to leading order we may use $\epsilon_\mrm{ep}$ or $\epsilon$
interchangeably.

The relation between Legendre polynomials in the
mid-point and end-point parametrisations is at leading order (for $\ell\geq1$)
% $\mathcal{L}_0(\mu_\mrm{ep})=\mathcal{L}_0(\mu)$ for the monopole and
\bea
\mathcal{L}_\ell(\mu)
=\mathcal{L}_\ell\big(\mu_\mrm{ep}+\tfrac12\epsilon\ssp(\mu_\mrm{ep}^2-1)\big)
&\simeq\mathcal{L}_\ell(\mu_\mrm{ep})+\frac\epsilon2\ssp(\mu_\mrm{ep}^2-1)
    \frac{\dif\mathcal{L}_\ell}{\dif\mu_\mrm{ep}}
=\mathcal{L}_\ell(\mu_\mrm{ep})
    +\frac\epsilon2\ssp\ell\ssp
    \big(\mu_\mrm{ep}\mathcal{L}_\ell(\mu_\mrm{ep})-\mathcal{L}_{\ell-1}(\mu_\mrm{ep})\big) \, ,
    \nonumber
\eea
where in the last equality we have used the recursion relation
$(x^2-1)\sp{\dif\mathcal{L}_\ell}/{\dif x}
=\ell\ssp\big(x\mathcal{L}_\ell(x)-\mathcal{L}_{\ell-1}(x)\big)$.
Thus, at leading order in $\epsilon$ we have
\be\nonumber
\xi_s(s,\mu)
\equiv\sum_{\ell}\xi^{(0)}_\ell(s)\mathcal{L}_\ell(\mu)
=\sum_{\ell}\xi^{(0)}_\ell(s)
    \Big[\mathcal{L}_\ell(\mu_\mrm{ep})
        +\frac\epsilon2\ssp\ell\ssp
    \big(\mu_\mrm{ep}\mathcal{L}_\ell(\mu_\mrm{ep})-\mathcal{L}_{\ell-1}(\mu_\mrm{ep})\big)
    \Big] \, .
\ee
Therefore, the multipoles in the mid-point parametrisation are related to those
in the end-point parametrisation by
\bea\label{eq:xil-ep}
\xi_\ell^\mrm{ep}(s)
&\equiv\frac{2\ell+1}{2}\int^1_{-1}\dif\mu_\mrm{ep}\,\mathcal{L}_{\ell}(\mu_\mrm{ep})\,
\xi_s(s,\mu)
=\xi^{(0)}_\ell(s)+\epsilon\sum_{\ell'}M_{\ell\ell'}\,\xi^{(0)}_{\ell'}(s)
    +\mathcal{O}(\epsilon^2) \, ,
\eea
where the coupling coefficients are given by
\be
M_{\ell\ell'}
\equiv\frac12\frac{2\ell+1}{2}\int^1_{-1}
    \dif\mu_\mrm{ep}\,\mathcal{L}_{\ell}(\mu_\mrm{ep})\ell'
    \big(\mu_\mrm{ep}\mathcal{L}_{\ell'}(\mu_\mrm{ep})-\mathcal{L}_{\ell'-1}(\mu_\mrm{ep})\big) \, .
\ee
The nonzero coefficients are $M_{12}=-3/5$, $M_{32}=3/5$, and
$M_{34}=-10/9$, i.e.~the only induced multipoles (at order $\epsilon$)
are for $\ell=1,3$, the dipole and octupole:
\be\nonumber
\xi^{\mrm{ep}(1)}_1=-\frac35\,\xi^{(0)}_2 \, ,
\qquad
\xi^{\mrm{ep}(1)}_3=\frac35\,\xi^{(0)}_2-\frac{10}{9}\,\xi^{(0)}_4 \, ,
\ee
i.e.\ there is a leakage of the even multipoles into
the odd multipoles. This expression agrees with
% equations 2.14 and 2.15 in Ref.~\cite{Beutler2018:1810.05051}.
equation 4.14 in Ref.~\cite{Reimberg_2016} upon inserting the
Kaiser multipoles~\eqref{eq:kaiser-multipoles}.

\bibliographystyle{Bonvinetal}
\bibliography{main}%,bibliography_drh}

\end{document}